\documentclass[twocolumn,
	aps, prd,
	10pt, notitlepage, 
        floats, floatfix,
	amsmath, amssymb, amsfonts, eqsecnum,
	superscriptaddress,
	showpacs, showkeys,
	nofootinbib,
 	longbibliography,
]{revtex4-1}

\usepackage{mathtools}
\usepackage{wrapfig}
\usepackage{graphicx} 
\usepackage[dvipsnames,table]{xcolor}
\usepackage{xspace} 
\usepackage{bm} 
\usepackage{amsmath,amsfonts,amssymb,amsthm,mathtools}
\usepackage[utf8]{inputenc} 
\usepackage{multirow}
\usepackage{array}

\xdefinecolor{mylinkcolor}{rgb}{0,0,0.5}
\usepackage[
	bookmarksnumbered, bookmarksopen, bookmarksopenlevel=2,
	breaklinks=true,
	colorlinks=true, filecolor=mylinkcolor, citecolor=mylinkcolor,
	linkcolor=mylinkcolor, urlcolor=mylinkcolor, menucolor=mylinkcolor,
]{hyperref}

\usepackage{verbatim}
\usepackage{mathrsfs}
\usepackage{color}
\usepackage{enumitem}
\usepackage{comment}
\usepackage{tikz}
\usepackage{tikz-3dplot}
\usepackage{mathbbol}
\usepackage{bbm}
\usepackage{siunitx}
\usetikzlibrary{decorations.pathmorphing}
\usetikzlibrary{arrows}

\tikzset{snake it/.style={decorate, decoration=snake}}

\usepackage[normalem]{ulem}

\newcommand{\be}{\begin{equation}}
\newcommand{\ee}{\end{equation}}
\newcommand{\bse}{\begin{subequations}}
\newcommand{\ese}{\end{subequations}}

\newcommand{\bpm}{\begin{pmatrix}}
\newcommand{\epm}{\end{pmatrix}}

\def\h{\mathbb{h}}
\def\A{\mathcal{A}}

\usepackage{colortbl}
\definecolor{blue2}{cmyk}{1, 0.1, 0.1, 0}

\definecolor{pyBlue}{RGB}{31, 119, 180}
\definecolor{pyRed}{RGB}{214, 39, 40}
\definecolor{pyGreen}{RGB}{44, 160, 44}
\definecolor{pyBlue2}{RGB}{0, 111, 237}
\definecolor{pyRed2}{RGB}{224, 52, 36}

\usepackage{colortbl}
\definecolor{summersky}{cmyk}{0.71,0.33,0,0.5}
\definecolor{flamingo}{cmyk}{0,0.51,0.71,0.5}
\definecolor{rp}{cmyk}{0.2, 1, 0.6, 0}
\definecolor{pacificblue}{cmyk}{0.95,0.3,0, 0.5}
\definecolor{gray60}{cmyk}{0.4,0.4,0,0.8}

\newcommand{\dd}{\mathop{\mathrm{d}\!}{}}

\begin{document}
\title{Searching for precessing binary systems with mode-by-mode filtering and marginalization}
\author{Zihan Zhou}
\email{zihanz@princeton.edu}
\affiliation{Department of Physics, Princeton University, Princeton, NJ 08540, USA}%

\author{Digvijay Wadekar}
\affiliation{\mbox{Weinberg Institute, University of Texas at Austin, Austin, TX 78712, USA}}
\affiliation{Department of Physics and Astronomy, Johns Hopkins University, 3400 N. Charles Street, Baltimore, Maryland, 21218, USA}

\author{Javier Roulet}
\affiliation{Kavli Institute for Cosmological Physics, The University of Chicago, 5640 South Ellis Avenue, Chicago, Illinois 60637, USA}
\affiliation{School of Natural Sciences, Institute for Advanced Study, 1 Einstein Drive, Princeton, NJ 08540, USA}

\author{Oryna Ivashtenko}
\affiliation{Department of Particle Physics and Astrophysics, Weizmann Institute of Science, Rehovot 7610001, Israel}

\author{Tejaswi Venumadhav}
\affiliation{\mbox{Department of Physics, University of California at Santa Barbara, Santa Barbara, CA 93106, USA}}
\affiliation{\mbox{International Centre for Theoretical Sciences, Tata Institute of Fundamental Research, Bangalore 560089, India}}

\author{Tousif Islam}
\affiliation{Kavli Institute for Theoretical Physics,University of California Santa Barbara, Kohn Hall, Lagoon Rd, Santa Barbara, CA 93106}

\author{Ajit Kumar Mehta}
\affiliation{\mbox{Chennai Mathematical Institute, Siruseri 603013, Chennai, India}}

\author{Jonathan Mushkin}
\affiliation{Department of Particle Physics and Astrophysics, Weizmann Institute of Science, Rehovot 7610001, Israel}

\author{Mark Ho-Yeuk Cheung}
\affiliation{\mbox{School of Natural Sciences, Institute for Advanced Study, 1 Einstein Drive, Princeton, NJ 08540, USA}}

\author{Barak Zackay}
\affiliation{Department of Particle Physics and Astrophysics, Weizmann Institute of Science, Rehovot 7610001, Israel}

\author{Matias Zaldarriaga}
\affiliation{\mbox{School of Natural Sciences, Institute for Advanced Study, 1 Einstein Drive, Princeton, NJ 08540, USA}}

\begin{abstract}
Nearly all previous binary black hole searches in LIGO–Virgo–KAGRA (LVK) gravitational wave data have assumed that the component spins are aligned with the orbital angular momentum, thereby neglecting spin-precession effects in the waveform, which can lead to potentially missing interesting signals. 
Precessing searches are challenging, because the extra degrees of freedom due to misaligned spins lead to: $(i)$ a much larger number of templates compared to the aligned-spin configurations, $(ii)$ an increased rate of background triggers.
To address this, we develop novel precessing signal template banks using mode-by-mode filtering and marginalization methods. We use the precession harmonic decomposition from~\citet{Fairhurst:2019vut} and filter each precessing harmonic separately with the data. We then marginalize over the SNRs from different harmonics in our detection statistic. We also use machine learning methods to improve our search efficiency: 
$(i)$ we use singular value decomposition together with random forest regressor to reduce redundancy in the dominant precessing-harmonic templates;
$(ii)$ we use normalizing flows to generate optimal prior samples for harmonic SNRs for the marginalized statistic.
We show that marginalizing (instead of maximizing) over the harmonic mode SNRs increases the search sensitive volume by $\sim 10\%$.
Results from searching in LVK data using this framework will be reported in a companion paper.
\end{abstract}

\maketitle

\section{Introduction}

Over the past decade, gravitational-wave (GW) observations by the LIGO/Virgo/KAGRA (LVK) collaboration have opened a new window onto the Universe \cite{LIGOScientific:2014pky,VIRGO:2014yos,KAGRA:2020agh}. To date, the LVK collaboration has reported more than 200 compact binary coalescences \cite{LIGOScientific:2020ibl,LIGOScientific:2021usb,KAGRA:2021vkt,LIGOScientific:2025slb}. The public release of LVK data through the Gravitational Wave Open Science Center has also enabled external groups to search for new events and identify additional candidates \cite{Nitz:2018imz,Venumadhav:2019tad,Venumadhav:2019lyq,Nitz:2020oeq,Nitz:2021uxj,Olsen:2022pin,Wadekar:2023gea,Koloniari2025}. Most detections have relied on template-based (matched-filter) searches, which correlate the detector's strain data with modeled waveforms \cite{Wadekar:2024zdq,LIGOScientific:2016dsl,LIGOScientific:2018mvr,Sachdev:2019vvd,Usman:2015kfa,2022CQGra..39e5002A,LIGOScientific:2021usb,KAGRA:2021vkt,LIGOScientific:2024elc,Venumadhav:2019lyq,Olsen:2022pin,Mehta:2023zlk,Chia:2023tle,Nitz:2020oeq,Nitz:2021uxj,Nitz:2021zwj,Cheung:2025grp}. These searches require template banks that are accurate and densely spaced to efficiently cover the targeted parameter space of compact-object mergers.

GW signals, as formally described by the multipole expansion in General Relativity (GR) \cite{Thorne:1980ru}, can be expressed as a sum over spin-weighted spherical harmonics. In typical cases, the quadrupole $(\ell=2, |m|=2)$ mode dominates the signal in the post-Newtonian (PN) regime \cite{Blanchet:2013haa}. However, higher-order modes—such as $(3,3)$, $(4,4)$, $(2,1)$, $(3,2)$, and $(4,3)$—grow increasingly important near the merger and ringdown phases, especially for binaries with unequal masses or those observed at high inclination angles \cite{Varma:2014jxa, Meh25_Population_HM,Wadekar:2023kym,Wadekar:2024zdq,Cheung:2025grp}. In addition to typically neglecting these higher-order modes, most existing template banks further assume the aligned-spin, quasi-circular approximation; that is, the component spins are taken to be parallel to the orbital angular momentum and the orbits nearly circular. While this scenario is expected for binaries formed via isolated binary evolution, dynamical formation channels in dense stellar environments can instead produce systems with nearly isotropic spin orientations \cite{Mandel:2018hfr,Franciolini:2022iaa,Mapelli:2021taw,Gerosa:2021mno,Kalogera:1999tq,Gerosa:2018wbw,Rodriguez:2016vmx}. Thus, observational measurements of the fraction of mergers exhibiting strong spin–orbit misalignment provide a crucial probe of the relative importance of different binary formation channels \cite{Stevenson:2017dlk,Gompertz:2021xub}. Ignoring precession in this case leads to the loss of sensitivity to such events (see Fig.~\ref{fig:volume_loss_precession_harmonics} below).

In general, searches for spin-precessing binaries are challenging because incorporating spin effects significantly increases the number of templates required. For example, Ref.~\cite{Schmidt:2024jbp} expands the physical parameter space by introducing two additional dimensions—the inclination angle and the precession opening angle—relative to standard aligned-spin searches. This approach substantially enlarges the template bank compared to the aligned-spin case and leads to a much greater computational cost. Similar challenges are encountered in searches aiming to account for higher-order mode contributions to the waveform \cite{Cha22, Wadekar:2023gea,Wadekar:2023kym,Wadekar:2024zdq, Meh25_VT_HM}.

To address these challenges, in this paper, we first construct a template bank based on the precession harmonics as described in Refs.~\cite{Fairhurst:2019vut,McIsaac:2023ijd,Pratten:2020ceb}. We then estimate the detection statistics using the mode-by-mode filtering and marginalization algorithms developed in Refs.~\cite{Wadekar:2023kym,Wadekar:2024zdq}. The key aspect of precession is encoded in the rotation between the orbital angular momentum $\boldsymbol{L}$ frame and the total angular momentum $\boldsymbol{J}$ frame, which can be straightforwardly computed using Wigner d-matrices. For any given multipolar sector $\ell$, the different $m$ modes in these two frames mix with one another. As a result, the instantaneous waveform in the $L$-frame can be expressed as a linear combination of harmonics in the $J$-frame. This harmonic decomposition motivates the use of a mode-by-mode matched-filtering strategy. This strategy was applied in Refs.~\cite{Wadekar:2023kym,Wadekar:2023gea} for a binary black hole search by decomposing aligned-spin GW waveforms as a sum of quasi-circular higher harmonics. Ref.~\cite{McIsaac:2023ijd} used this strategy for a precessing neutron-star black hole search. Recently, Ref.~\cite{Dhurkunde:2026jhp} searched for precessing binaries in a limited mass range and for highly asymmetric mass ratios ($q = m_2/m_1 < 0.2$). In this paper, we focus on the complementary regime: systems with $q > 0.2$ covering a broader mass range. In practice, we start by building template banks for the dominant precession harmonic ($k=0$), then associate each with the four subdominant harmonics ($k=1,2,3,4$). To organize these samples, we use the \texttt{KMeans} clustering algorithm. For phase modeling in each bank, we combine singular value decomposition (SVD) with a machine-learning algorithm called random forest (RF) regressor. This combined method achieves matches greater than $90\%$ for more than $99\%$ of waveforms in our training set.

After independently filtering each precession harmonic, we combine the five resulting SNR time series and marginalize over the binary inclination and sky location. Unlike methods which explicitly sample over these extrinsic parameters, the mode-by-mode approach achieves this with only a modest increase in computational cost compared to a standard aligned-spin search. In particular, when combining the five precession harmonics, we evaluate the full likelihood by marginalizing over the relative SNR ratios, as opposed to simply maximizing over them. This strategy effectively suppresses noise triggers, reduces the false-alarm rate, and thus improves the search sensitivity. Preliminary results indicate that marginalization can yield an increase of approximately $10\%$ in the sensitive volume.

The remainder of this paper is organized as follows. In Section~\ref{sec:waveform}, we review the essential properties of gravitational waves from precessing binaries and outline how precession harmonics are constructed. Section~\ref{sec:bank} discusses our modeling of template-bank amplitudes and phases. In Section~\ref{sec:marginalize}, we present the matched-filtering framework, highlighting the construction of the marginalized likelihood. The effectualness of the template bank is assessed in Section~\ref{sec:effectualness}. Finally, in Section~\ref{sec:con_future}, we summarize our results and discuss potential improvements and future extensions.

Throughout the paper, we adopt the following commonly used notations: $m_1, m_2$ for the individual black hole masses. $\boldsymbol{S}_1,\boldsymbol{S}_2$ are their spins. The following combined variables are typically used in the waveform models:
\begin{equation}
\begin{aligned}
M & \equiv m_1+m_2 ~, \, \eta \equiv m_1 m_2 / M^2 ~, \, q=m_2/m_1 ~,  \\
\delta & \equiv \left(m_1-m_2\right) / M ~, \, \chi_i\equiv \mathbf{S}_i / m_i^2~, \\
\boldsymbol{\chi}_s & \equiv \left(\boldsymbol{\chi}_1+\boldsymbol{\chi}_2\right) / 2 ~, \, \boldsymbol{\chi}_a \equiv \left(\boldsymbol{\chi}_1-\boldsymbol{\chi}_2\right) / 2 ~.
\end{aligned}
\end{equation}
We use the following definition for the noise-weighted inner product 
\begin{equation}
\left\langle h_i \mid h_j\right\rangle=4 \int_0^{\infty} \mathrm{d} f \frac{h_i^*(f) h_j(f)}{S_n(f)} ~,
\end{equation}
where $S_n(f)$ is the one-sided power spectral density
(PSD). 

\begin{figure*}[t]
    \centering
    \includegraphics[width=0.25\textwidth]{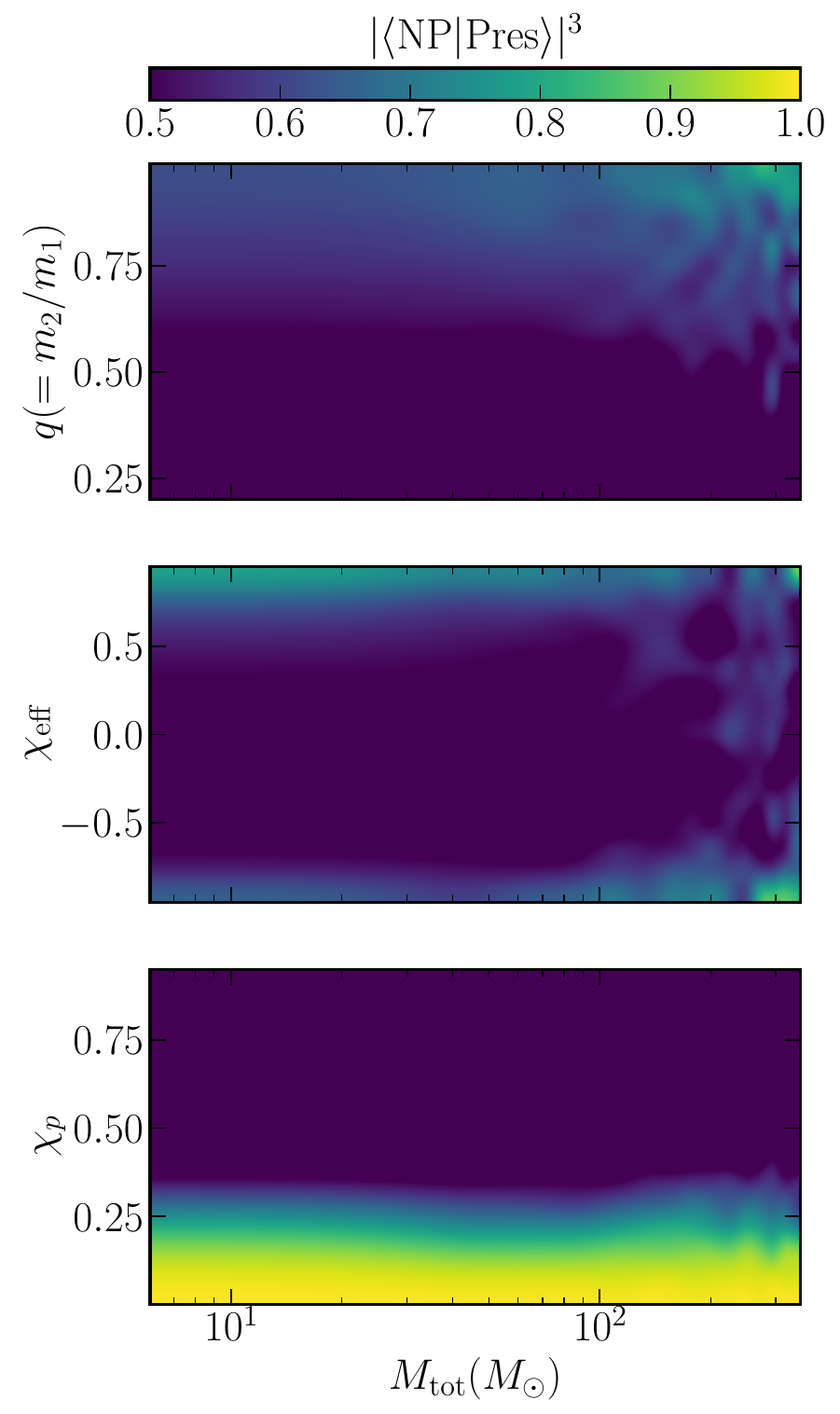}\hfill
    \includegraphics[width=0.25\textwidth]{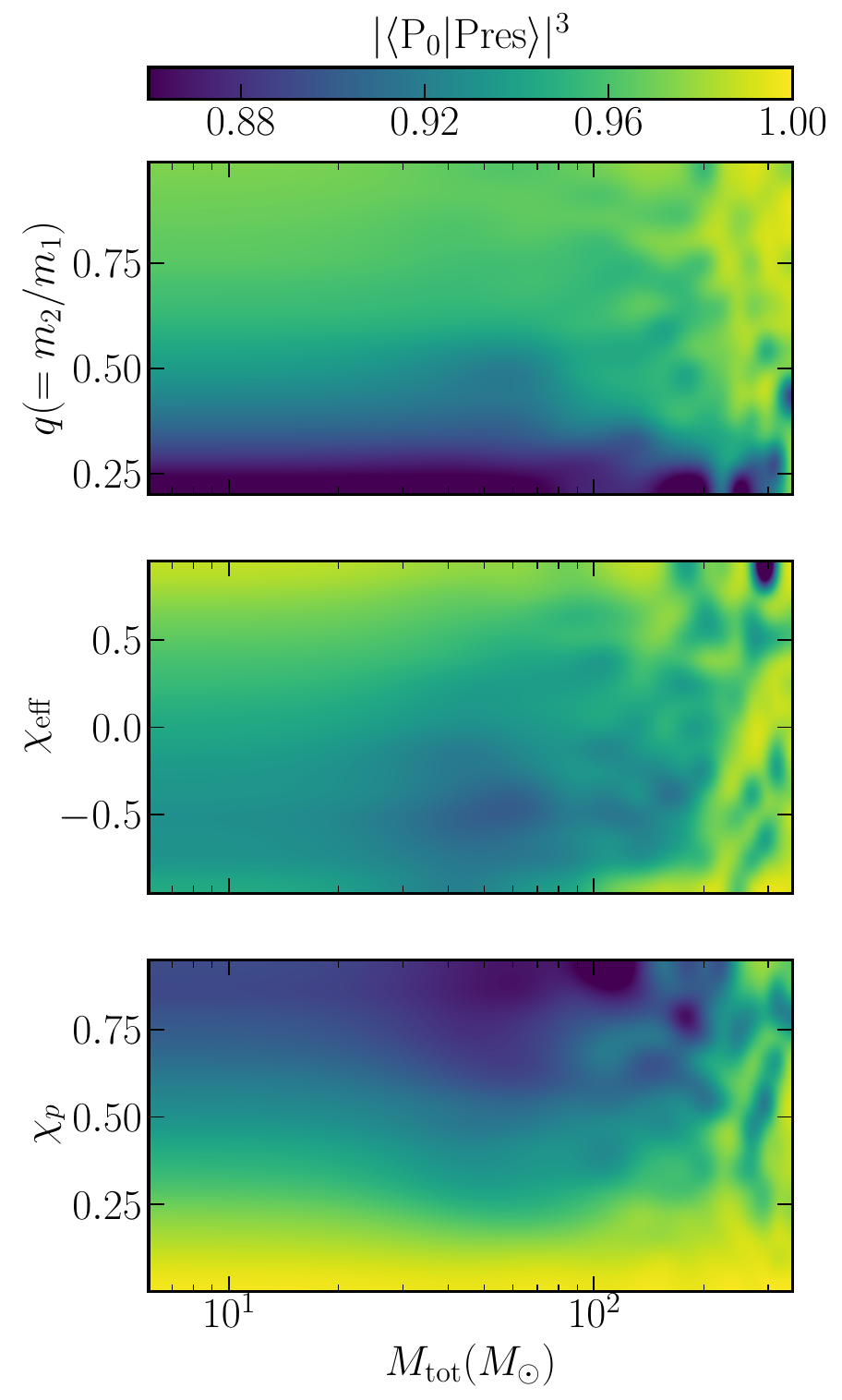}\hfill
    \includegraphics[width=0.25\textwidth]{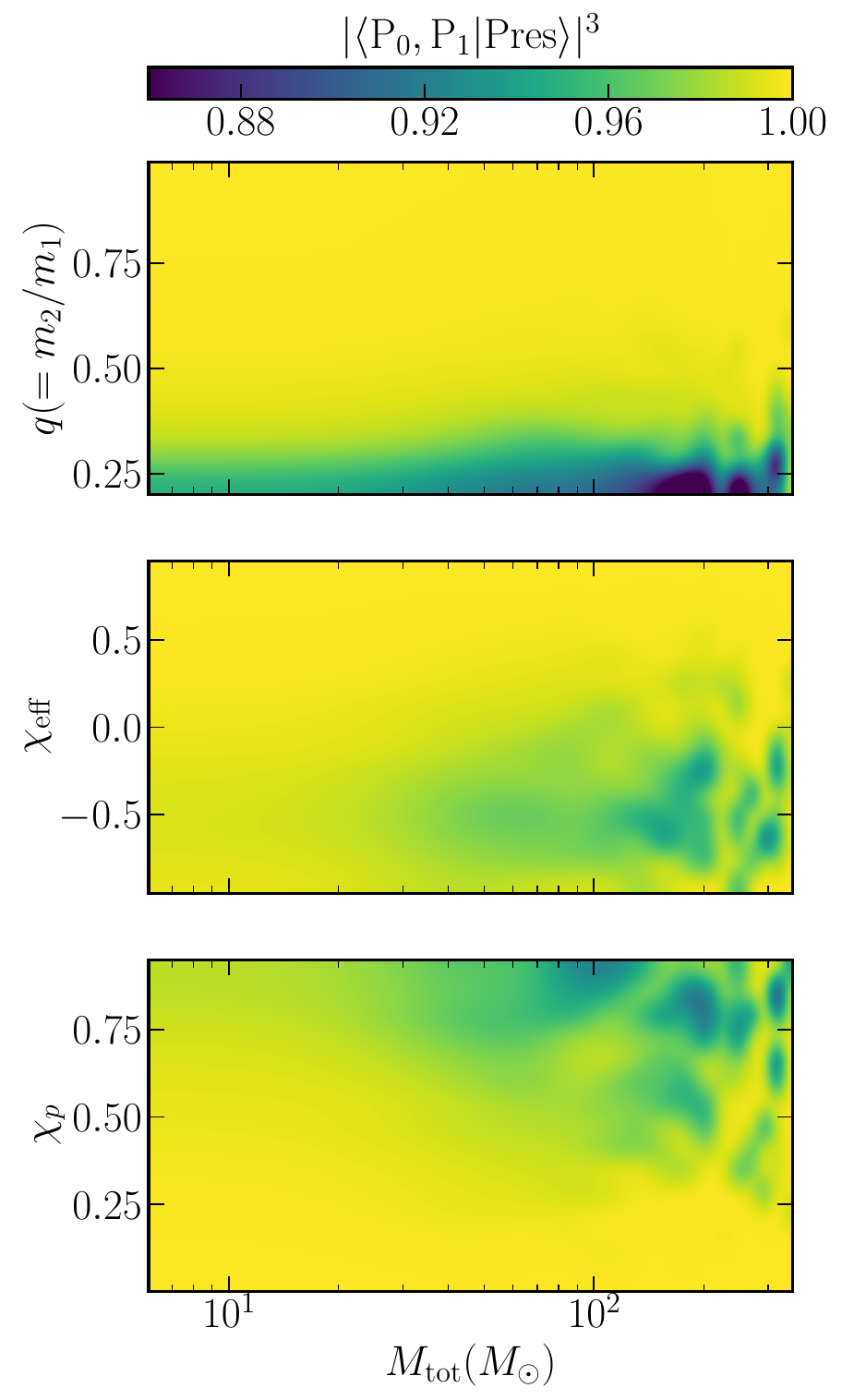}\hfill
    \includegraphics[width=0.25\textwidth]
    {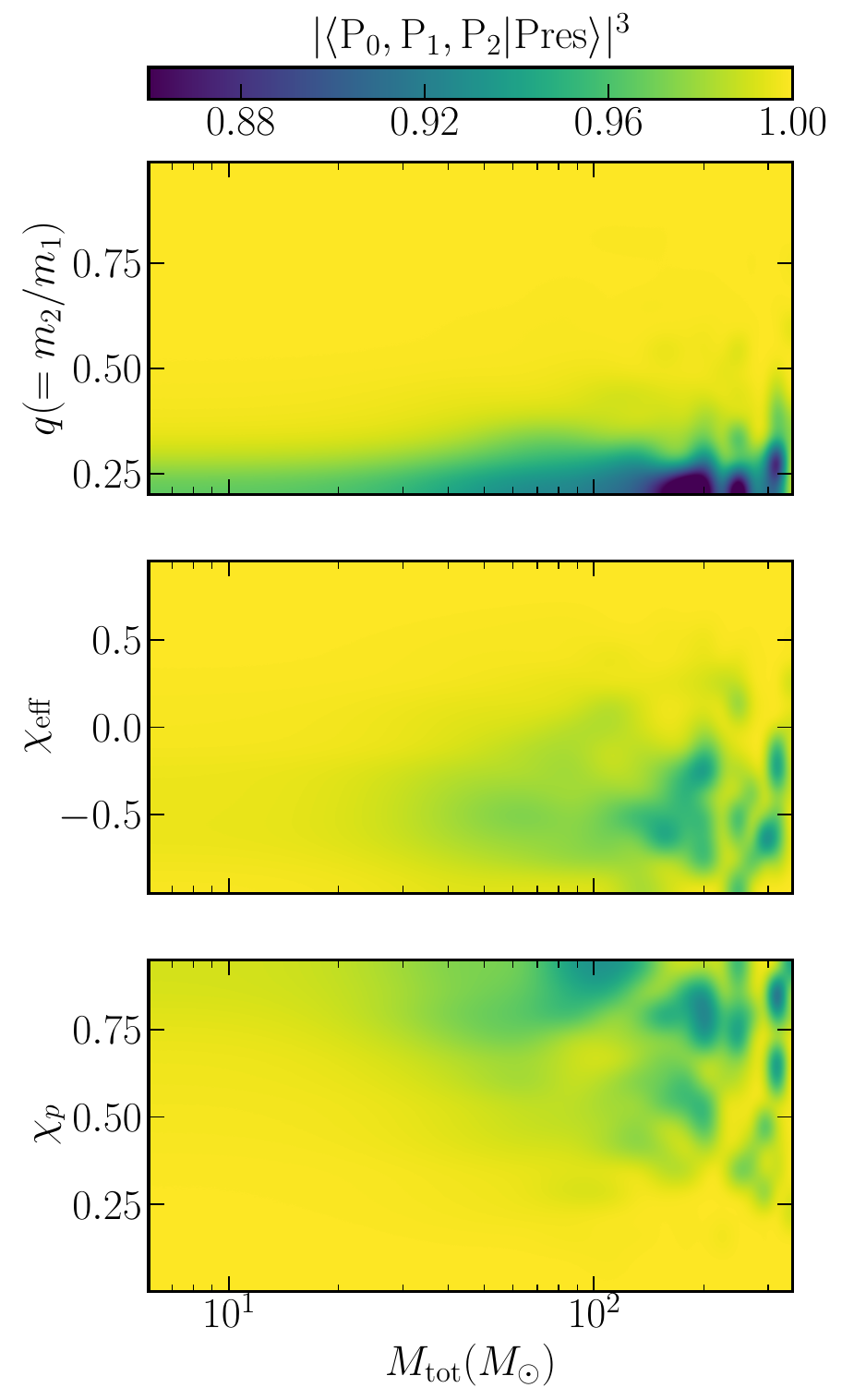}
    \caption{
    Neglecting precession harmonics in the template waveforms of a matched-filtering search can reduce the overlap with a fully precessing signal and thereby decrease the sensitive (detection) volume.
    In this figure we quantify this effect across the parameter space shown by plotting the fractional volume loss (calculated by the cube of the waveform overlap in this simplistic scenario).
    The four panels correspond to increasingly complete the precession harmonics in the templates:
    nonprecessing (NP), and precession-harmonics
    ${\rm P}_0$, $\rm P_1$, and $\rm P_2$.
    As more harmonics are included, the mismatch-driven loss is progressively reduced. For mass-ratio $q>0.2$, the first two harmonics play the important role while the third one only gives marginal improvement. 
    }
    \label{fig:volume_loss_precession_harmonics}
\end{figure*}

\section{Constructing Precession Harmonic Waveforms}
\label{sec:waveform}

In this section, we briefly review the construction of precession-harmonic waveforms. Much of this material is well established in the literature; for comprehensive reviews and technical details, we refer the reader to Refs.~\cite{Farr:2014qka,Fairhurst:2019vut,Pratten:2020ceb,McIsaac:2023ijd}.

\subsection{Binary Configuration}

In general relativity, a binary system consists of two objects with masses $m_1$ and $m_2$ (with $m_1 \geq m_2$, so the mass ratio $q = m_2/m_1 \leq 1$) and spins $\boldsymbol{S}_1 = m_1^2 \boldsymbol{\chi}_1$ and $\boldsymbol{S}_2 = m_2^2 \boldsymbol{\chi}_2$. During the inspiral phase, the two objects orbit one another and gradually spiral together as gravitational waves carry away energy and angular momentum. If the spins and the orbital angular momentum $\boldsymbol{L}$ are aligned, the orbital plane remains fixed. However, if the spins are misaligned with respect to $\boldsymbol{L}$, the orbital plane does not stay fixed and instead precesses around the total angular momentum $\boldsymbol{J} = \boldsymbol{L} + \boldsymbol{S}_1 + \boldsymbol{S}_2$.

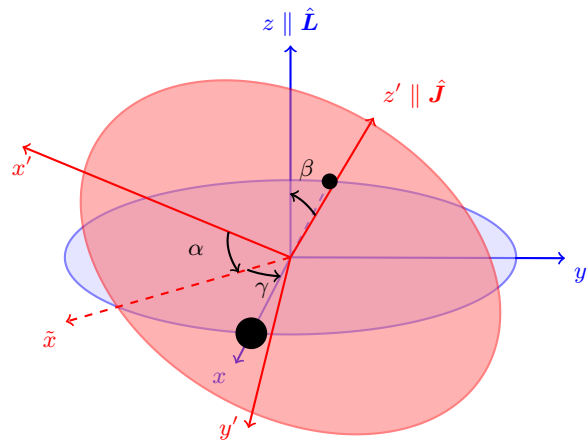
\begin{figure}[h!]
    \centering
    \tdplotsetmaincoords{70}{100}
    \begin{tikzpicture}[tdplot_main_coords, scale=3]

  \coordinate (O) at (0,0,0);

  %
  %
  %
  %
  %

  \def\betaA{35}     
  \def\phiJ{50}      
  \def\alphaA{45}    

  \draw[->, thick, blue]   (O) -- (1.4,0,0)   node[anchor=north east] {$x$};
  \draw[->, thick, blue]   (O) -- (-0.2,1.2,0) node[anchor=north west] {$y$};
  \draw[->, thick, blue]   (O) -- (0,0,1.0)   node[anchor=south]      {$z\parallel \hat{\boldsymbol{L}}$};
  \draw[dashed, thick, blue] (O) -- (-1.0,0,0);

  \draw[thick, draw=blue, fill=blue!20, opacity=0.5]
    plot[smooth,domain=0:360,samples=120,variable=\t]
      ({cos(\t)},{sin(\t)},0) -- cycle;

  \draw[thick, draw=red, fill=red!60, opacity=0.5]
    plot[smooth,domain=0:360,samples=200,variable=\t]
      ({sin(\phiJ)*cos(\t) + cos(\betaA)*cos(\phiJ)*sin(\t)},
       {-cos(\phiJ)*cos(\t) + cos(\betaA)*sin(\phiJ)*sin(\t)},
       {-sin(\betaA)*sin(\t)})
    -- cycle;

  \coordinate (zJ) at
    ({sin(\betaA)*cos(\phiJ)},
     {sin(\betaA)*sin(\phiJ)},
     {cos(\betaA)});
  \draw[->, thick, red] (O) -- (zJ)
    node[anchor=south west] {$z' \parallel \hat{\boldsymbol{J}}$};

  \coordinate (xtilde) at ({1.3*sin(\phiJ)}, {-1.3*cos(\phiJ)}, 0);
  \draw[->, dashed, thick, red] (O) -- (xtilde)
    node[anchor=north east] {$\tilde{x}$};

  \coordinate (xprime) at (
    {1.3*(sin(\phiJ)*cos(\alphaA) - cos(\betaA)*cos(\phiJ)*sin(\alphaA))},
    {1.3*(-cos(\phiJ)*cos(\alphaA) - cos(\betaA)*sin(\phiJ)*sin(\alphaA))},
    {1.3*(sin(\betaA)*sin(\alphaA))}
  );
  \draw[->, thick, red] (O) -- (xprime) node[anchor=north] {$x'$};

  \coordinate (yprime) at (
    {1.1*(sin(\phiJ)*sin(\alphaA) + cos(\betaA)*cos(\phiJ)*cos(\alphaA))},
    {1.1*(-cos(\phiJ)*sin(\alphaA) + cos(\betaA)*sin(\phiJ)*cos(\alphaA))},
    {1.1*(-sin(\betaA)*cos(\alphaA))}
  );
  \draw[->, thick, red] (O) -- (yprime) node[anchor=east] {$y'$};


  \draw[->, thick, black]
    plot[smooth,domain=-\alphaA:0,samples=50,variable=\s]
      ({0.3*(sin(\phiJ)*cos(\s) + cos(\betaA)*cos(\phiJ)*sin(\s))},
       {0.3*(-cos(\phiJ)*cos(\s) + cos(\betaA)*sin(\phiJ)*sin(\s))},
       {0.3*(-sin(\betaA)*sin(\s))});
  \coordinate (Amid) at
    ({0.38*(sin(\phiJ)*cos(-\alphaA/2) + cos(\betaA)*cos(\phiJ)*sin(-\alphaA/2))},
     {0.38*(-cos(\phiJ)*cos(-\alphaA/2) + cos(\betaA)*sin(\phiJ)*sin(-\alphaA/2))},
     {0.38*(-sin(\betaA)*sin(-\alphaA/2))});
  \node at (Amid) [anchor=east] {$\alpha$};

  \draw[->, thick, black]
    plot[smooth,domain=0:\betaA,samples=50,variable=\s]
      ({0.3*(sin(\betaA)*cos(\phiJ)*cos(\s) - cos(\betaA)*cos(\phiJ)*sin(\s))},
       {0.3*(sin(\betaA)*sin(\phiJ)*cos(\s) - cos(\betaA)*sin(\phiJ)*sin(\s))},
       {0.3*(cos(\betaA)*cos(\s) + sin(\betaA)*sin(\s))});
  \coordinate (Bmid) at
    ({0.37*(sin(\betaA)*cos(\phiJ)*cos(\betaA/2) - cos(\betaA)*cos(\phiJ)*sin(\betaA/2))},
     {0.37*(sin(\betaA)*sin(\phiJ)*cos(\betaA/2) - cos(\betaA)*sin(\phiJ)*sin(\betaA/2))},
     {0.37*(cos(\betaA)*cos(\betaA/2) + sin(\betaA)*sin(\betaA/2))});
  \node at (Bmid) [anchor=south] {$\beta$};

  \tdplotdrawarc[
    ->, thick, black
  ]{(O)}{0.25}{\phiJ-90}{0}{below}{$\gamma$}

  \coordinate (P1) at (1,0,0);
  \coordinate (P2) at (-1,0,0);
  \fill (P1) circle (2pt);
  \fill (P2) circle (1pt);

\end{tikzpicture}
    \caption{The configuration of a binary black hole system with orbital angular momentum $\boldsymbol{L}$ and total angular momentum $\boldsymbol{J}$ misaligned. In the $L$-frame (blue), $\hat{\boldsymbol{z}} \parallel \hat{\boldsymbol{L}}$, and $\hat{\boldsymbol{x}}$ is defined as the unit vector pointing from $m_2$ to $m_1$. The $J$-frame (red) is defined with $\hat{\boldsymbol{z}}' \parallel \hat{\boldsymbol{J}}$, and the line-of-sight vector $\hat{\boldsymbol{N}}$ lies in the $x'-z'$ plane. The three Euler angles $(\alpha,\beta,\gamma=-\epsilon)$ describe the active rotation from the $J$-frame to the $L$-frame in the $(z,y,z)$ convention~\cite{Pratten:2020ceb}. Here $\tilde x$ denotes the line of nodes (intersection of the two equatorial planes), $\alpha$ is measured in the $J$-frame equatorial plane, $\beta$ is the opening angle between $\hat{\boldsymbol{J}}$ and $\hat{\boldsymbol{L}}$, and $\gamma$ is measured in the $L$-frame equatorial plane.}
    \label{fig:BBH_config}
\end{figure}

We draw the binary configuration for precessing system in Fig.~\ref{fig:BBH_config}, 
where $\alpha,\beta$ and $\epsilon$ are three Euler angles that capture the frame rotation from $\boldsymbol{L}$ to $\boldsymbol{J}$. We follow the LAL conventions of the coordinate system described in Refs.~\cite{Pratten:2020ceb,Yu:2023lml}, in particular Appendix C of Ref.~\cite{Pratten:2020ceb}. The $L$-frame (co-precessing frame) is defined such that the $z$-axis is aligned with the instantaneous orbital angular momentum $\boldsymbol{L}$, while the $x$-axis points from the secondary mass $m_2$ toward the primary mass $m_1$. In this non-inertial frame, the binary orbital plane remains fixed, and the effects of precession appear as a rotation of the frame relative to the inertial observer. In this case, the line-of-sight unit vector $\hat{\boldsymbol{N}}$ is given by
\begin{equation}
\label{eq:line_of_sight}
    \hat{\boldsymbol{N}}\equiv\left(\begin{array}{c}
\sin \iota \cos \left(\pi / 2-\phi_{\mathrm{orb}}\right) \\
\sin \iota \sin \left(\pi / 2-\phi_{\mathrm{orb}}\right) \\
\cos \iota
\end{array}\right)_{L_0} ~,
\end{equation}
where $L_0$ denotes the co-precessing frame at $f=f_{\rm ref}$. $\iota$ is the inclination angle and $\phi_{\rm orb}$ is the orbital phase at the reference frequency $f_{\rm ref}$. Moreover, we can straightforwardly read off the total angular momentum in the $L$-frame
\begin{equation}
\label{eq:J_eq}
    \boldsymbol{J} \equiv \left(\begin{array}{c}
m_1^2 \chi_{1 x}+m_2^2 \chi_{2 x} \\
m_1^2 \chi_{1 y}+m_2^2 \chi_{2 y} \\
L+m_1^2 \chi_{1 z}+m_2^2 \chi_{2 z}
\end{array}\right)_L.
\end{equation}
With Eq.~\eqref{eq:line_of_sight} and Eq.~\eqref{eq:J_eq}, we can calculate the angles between $\boldsymbol{J},\boldsymbol{N}$
\begin{equation}
    \theta_{\mathrm{JN}}  \equiv \arccos \left(\frac{\boldsymbol{J} \cdot \hat{\boldsymbol{N}}}{|\boldsymbol{J}|}\right)
\end{equation}
and the precession opening angles between $\boldsymbol{L}$ and $\boldsymbol{J}$
\begin{equation}
    \begin{aligned}
 \beta  \equiv \theta_{\rm JL} = {\rm arccos} \frac{J_{z,L}}{|\boldsymbol{J}|} ~.
\end{aligned}
\end{equation}
The $J$-frame is chosen such that the $z'$-direction is aligned with $\boldsymbol{J}$, and the $\hat{\boldsymbol{N}}$ vector is in the $x'$-$z'$ plane
\begin{equation}
    \hat{\boldsymbol{N}} = \left(\begin{array}{c}
\sin \theta_{\rm J N} \\
0 \\
\cos \theta_{\rm J N}
\end{array}\right)_{J} ~.
\end{equation}
With the above frame definitions, the Euler angles $(\alpha,\beta,\gamma=-\epsilon)$ can be expressed entirely in terms of the vectors $\boldsymbol{J}$, $\boldsymbol{L}$, and $\hat{\boldsymbol{N}}$. The complete formulas are given in Appendix C of Ref.~\cite{Pratten:2020ceb}. In particular, the orbital angular momentum $\boldsymbol{L}$ in $J$-frame can be written as
\begin{equation}
    \hat{\boldsymbol{L}} = \left(\begin{array}{c}
\sin \beta \cos \alpha \\
\sin \beta \sin \alpha \\
\cos \beta
\end{array}\right)_{J} ~.
\end{equation}

Regarding the spin precession dynamics, $\boldsymbol{L},\boldsymbol{S}_1,\boldsymbol{S}_2$ and $\boldsymbol{J}$ evolve according to the spin precession equations (Eqs.~(10)–(12) in Ref.~\cite{Yu:2023lml} and Eqs.~(1a)–(1c) in Ref.~\cite{Akcay:2020qrj}). In the PN regime, gravitational wave emission changes $\boldsymbol{J}$ only on the radiation-reaction timescale, which is typically much longer than the precession period. As a result, the total angular momentum is quasi-conserved and can be regarded as inertial frame \footnote{More rigorously, we should choose the inertial frame $J_0$ that coincides with $J$-frame at $f=f_{\rm ref}$.}. The orbital angular momentum $\boldsymbol{L}$ precesses around $\boldsymbol{J}$, with its orbit-averaged value at 2.5PN order given by
\begin{equation}
\begin{aligned}
\boldsymbol{L}=\frac{\eta}{v} & \left\{\hat{\boldsymbol{L}}_{\mathrm{N}}\left[1+v^2\left(\frac{3}{2}+\frac{\eta}{6}\right)+v^4\left(\frac{27}{8}-\frac{19 \eta}{8}+\frac{\eta^2}{24}\right)\right]\right. \\
& \left.+v^3 \Delta \boldsymbol{L}_{1.5 \mathrm{PN}}^S+v^5 \Delta \boldsymbol{L}_{2.5 \mathrm{PN}}^S\right\} ~,
\end{aligned}
\end{equation}
where $\Delta \boldsymbol{L}_{\rm n PN}^S$ are spin corrections for which we take the orbital-averaged value from Ref.~\cite{Akcay:2020qrj}. The leading order PN expression for $\alpha(f)$ is given by \cite{Pratten:2020ceb}
\begin{equation}
\begin{aligned}
\label{eq:alpha_f_pn}
   \alpha(f) & = \Bigg(-\frac{5 (m_1 - m_2)}{64 m_1} - \frac{35}{192}\Bigg) (\pi f)^{-1}  + \cdots ~.
\end{aligned}
\end{equation}
The higher order terms of $\alpha(f)$ (and $\epsilon(f)$) can be found in Appendix G of Ref.~\cite{Pratten:2020ceb}.

\subsection{Harmonic Precession Waveforms}

We are now going to discuss the harmonic construction of the precessing signals as introduced in \cite{Pratten:2020ceb}. The essential idea is to rotate the instantaneous waveform in the co-precessing $L$-frame to get the waveform in the inertial $J$-frame. More explicitly, the total GW strain in the $J$-frame can be decomposed as
\begin{equation}
\label{eq:J_mode}
    h(t) = h_{+}^{J}-i h_{\times}^{J}=\sum_{\ell, m} h_{\ell m}^{J} \, {}_{-2}Y_{\ell m}\left(\theta_{\rm J N}, \phi_{\rm J N}\right) ~.
\end{equation}
In our convention, $\phi_{\rm JN}=0$ as $\hat{\boldsymbol{N}}$ is in the $x'$-$z'$ plane. The rotation from $L$-frame to $J$-frame indicates the following relation for the harmonic mode functions
\begin{equation}
    h_{\ell m}^{J}(t)=e^{-i m \alpha} \sum_{m^{\prime}} e^{i m^{\prime} \epsilon} d_{m m^{\prime}}^{\ell}(\beta) h_{\ell m^{\prime}}^L(t)
\end{equation}
where $d_{m m'}^\ell(\beta)$ denote the real-valued
Wigner d-matrices. Plugging the above expression into Eq.~\eqref{eq:J_mode}, we get
\vspace{-2\baselineskip}
\begin{widetext}
    \begin{align}
        h_+^J(t) & = \frac{1}{2} \sum_{\ell \geq 2} \sum_{m=-\ell}^\ell \Bigg[\sum_{m^{\prime}=-\ell}^{\ell} e^{i m^{\prime} \epsilon} e^{-i m \alpha} d_{m m^{\prime}}^{\ell}(\beta) {h}_{\ell m^{\prime}}^L(t)\,_{-2} Y_{\ell m}+\sum_{m^{\prime}=-\ell}^{\ell} e^{-i m^{\prime} \epsilon} e^{i m \alpha} d_{m m^{\prime}}^{\ell}(\beta) {h}_{\ell m^{\prime}}^{L *}(t)\,_{-2} Y_{\ell m}^*\Bigg] \\
        {h}_{\times}^J(t) & =\frac{i}{2} \sum_{\ell \geq 2} \sum_{m=-\ell}^{\ell}\left[\sum_{m^{\prime}=-\ell}^{\ell} e^{i m^{\prime} \epsilon} e^{-i m \alpha} d_{m m^{\prime}}^{\ell}(\beta) {h}_{\ell m^{\prime}}^L(t)\,_{-2} Y_{\ell m}-\sum_{m^{\prime}=-\ell}^{\ell} e^{-i m^{\prime} \epsilon} e^{i m \alpha} d_{m m^{\prime}}^{\ell}(\beta) {h}_{\ell m^{\prime}}^{L *}(t)\,_{-2} Y_{\ell m}^*\right]
    \end{align}
\end{widetext}
In the $L$-frame, the waveforms are essentially the instantaneous non-precessing ones. In this paper, we are mainly going to focus on the search of spin-precessing systems and ignore the contributions from the higher-order aligned-spin harmonics. This allows us to restrict our analysis to only the $|m'|=2$ sector. Once we plug in the amplitude / phase for the $L$-frame waveforms and include the detector response $F_+,F_\times$, we can write the whole GW strain in Eq.~\eqref{eq:J_mode} as 
\begin{equation}
\begin{aligned}
\label{eq:time_domain_waveform}
h(t)& =\frac{d_o}{D_L}\Re\Bigg\{\frac{A_0(t) e^{2 i\left(\phi_s(t)+\alpha(t)\right)}}{\left(1+b^2(t)\right)^2} \Bigg[\sum_{k=0}^4\left(b(t) e^{-i \alpha(t)}\right)^k \\
& \quad \times \left(F_{+} \mathcal{A}_k^{+}-i F_{\times} \mathcal{A}_k^{\times}\right)\Bigg]\Bigg\} ~,
\end{aligned}
\end{equation}
where the amplitude $A_0(t)$ is proportional to the non-precessing $(2,2)$ mode in $L$-frame. The parameter
\begin{equation}
    b\equiv  \tan(\beta/2)
\end{equation}
quantifies the precession opening angle between $\boldsymbol{J}$ and $\boldsymbol{L}$. $\Re$ denotes the real part of the above expression. The phase $\phi_s(t)$ is given by
\begin{equation}
    \phi_s(t) = \phi_{\rm orb}(t) - \epsilon(t) ~,
\end{equation}
with $\phi_{\rm orb}$ the orbital phase. The evolution of the third Euler angle $\epsilon(t)$ is obtained using the minimal rotation condition \cite{Pratten:2020ceb}
\begin{equation}
    \dot \epsilon \simeq \dot \alpha \cos\beta ~.
\end{equation}
In the precession harmonic decomposition, the inclination dependence of the system with respect to the line-of-sight vector $\hat{\boldsymbol{N}}$ is given by
\begin{widetext}
\begin{equation}
\begin{aligned}
\label{eq:A_angle}
\mathcal{A}_0^{+} = \mathcal{A}_4^{+} = \left( \frac{1 + \cos^2 \theta_{\mathrm{JN}}}{2} \right) ~, \\
\mathcal{A}_0^{\times} = -\mathcal{A}_4^{\times} = \cos \theta_{\mathrm{JN}} ~, \\
\mathcal{A}_1^{+} = -\mathcal{A}_3^{+} = 2 \sin \theta_{\mathrm{JN}} \cos \theta_{\mathrm{JN}} ~, \\
\mathcal{A}_1^{\times} = \mathcal{A}_3^{\times} = 2 \sin \theta_{\mathrm{JN}} ~, \\
\mathcal{A}_2^{+} = 3 \sin^2 \theta_{\mathrm{JN}} ~, \quad \mathcal{A}_2^{\times} = 0 ~.
\end{aligned}
\end{equation}
\end{widetext}
Here, $\theta_{\rm JN}$ can be treated as time independent constant thanks to the quasi-conservation of total angular momentum.

Transforming into the frequency domain under the stationary phase approximation (SPA) \cite{bender2013advanced}, the time domain waveform in Eq.~\eqref{eq:time_domain_waveform} can be written as
\begin{equation}
\begin{aligned}
\label{eq:harm_wave}
    h(f) = \frac{D_0}{D_L} \sum_{k=0}^4 [F_{+} \mathcal{A}_k^{+}-i F_{\times} \mathcal{A}_k^{\times}]\, R_k\, \h_k(f) ~,
\end{aligned}
\end{equation}
where $\h_k(f)$ is the  waveform corresponding to each harmonic. $R_k$ is the complex mode ratio for each mode ($R_0=1$). Note that we have absorbed all the frequency dependent terms in $\h_k(f)$ and the rest of the terms in Eq.~\eqref{eq:harm_wave} are complex scalars. This will be important for the marginalization procedure discussed below. The full expression for $\h_k(f)$ is given by
\begin{widetext}
\begin{equation}
\label{eq:pres_harm}
    \h_k(f) = \frac{A_0(f)}{D_k} \frac{b(f)^k}{(1+b(f)^2)^2} \exp\Big[2i \Big(\phi_s(f)-\phi_s(f_{\rm ref}) \Big) + i (2-k) \Big(\alpha(f)-\alpha(f_{\rm ref}) \Big)\Big] ~,
\end{equation}
\end{widetext}
where the proportionality constant $D_k$ is chosen such that each mode template is normalized according to a reference power spectral density as $\langle \h_k | \h_k \rangle = 1$. Note that $\h_k(f)$ are defined such that the phases of modes at a reference frequency $f_{\rm ref}$ are set to zero.
We will use the normalized mode waveforms $\h_k$ in section \ref{sec:bank} below to construct the mode templates used in our template banks. The relative phase between different modes at $f_{\rm ref}$ is absorbed into $R_k$ as
\begin{equation}
\begin{aligned}
    \arg(R_k) & = 2 \phi_s(f_{\rm ref}) + (2-k) \alpha(f_{\rm ref}) ~.
\end{aligned}
\end{equation}
It is worth highlighting here that the dominant mode $k=0$ is not the same as the non-precessing waveform $(2,2)$ because of the phase modulation from the first Euler angle $\alpha(f)$. The harmonic decomposition for the precession waveforms allows us to perform the search by first doing the mode-by-mode matched filtering and then marginalize over the mode amplitude ratio. Similar ideas have been used in Refs.~\cite{Wadekar:2023gea,Wadekar:2023kym,Wadekar:2024zdq,Cheung:2025grp} for the search of GW signals including higher aligned-spin harmonics, and proved to be more efficient by greatly reducing the computational cost.

Given that $\theta_{\rm JN}$ is almost time-independent, we can extract different harmonics by viewing the system from different $\theta_{\rm JN}$. More precisely, comparing the waveform in \texttt{IMRPhenomXPHM} \cite{Pratten:2020ceb}
\begin{equation}
    h^{\rm XPHM} (f) = F_+ h^{\rm XPHM}_+(f) + F_\times h_\times^{\rm XPHM}(f) ~,
\end{equation}
with the harmonic waveform decomposition in Eq.~\eqref{eq:harm_wave}, we get the dominant precession harmonics
\begin{equation}
    h_0 = \frac{1}{2} ( h_+^{\rm XPHM} + i h_\times^{\rm XPHM}) \Big|_{\theta_{\rm JN}=0} ~,
\end{equation}
and four associated higher harmonics
\begin{equation}
\begin{aligned}
    h_4 & = \frac{1}{2}(h_+^{\rm XPHM} - i  h_\times^{\rm XPHM}) \Big|_{\theta_{\rm JN}=0} ~, \\
    h_2  & = \frac{1}{3} \Big(h_+^{\rm XPHM} \Big|_{\theta_{\rm JN} = \pi/2} - \frac{1}{2} h_0 - \frac{1}{2} h_4\Big)  ~, \\
    h_1 & = \frac{1}{8}  \Big((2 i \sqrt{2} h_\times^{\rm XPHM} + 4  h_+^{\rm XPHM}) \Big|_{\theta_{\rm JN} = \pi/4} \\
    & \quad - 5 h_0 - 6  h_2 -  h_4 \Big) ~, \\
    h_3 & = \frac{1}{8} \Big( (2i \sqrt{2} h_\times^{\rm XPHM} - 4 h_+^{\rm XPHM}) \Big|_{\theta_{\rm JN} = \pi/4} \\
    & \quad + h_0 + 6  h_2 +5  h_4\Big) ~.
\end{aligned}
\end{equation} 
We then use these samples to generate a library of complex mode ratios $R_{k}$ samples using
\begin{equation}
    R_{k} = \frac{h_k(f_{\rm ref})}{h_0(f_{\rm ref})}
\label{eq:mode_ratio}
\end{equation}
and then compute the normalized mode waveforms $\h_k (f)$ as
\begin{equation}
    \h_k (f) = \frac{h_k(f)}{\langle h_k | h_k \rangle^{1/2}} e^{-i \arg(R_k)}.
\end{equation}
It can sometimes be convenient to perform Gram--Schmidt orthogonalization of different mode waveforms. We denote the orthogonalized mode waveforms by $h^\perp_k$ (we use the convention $h_{k=0} = h^\perp_{k=0}$) and the corresponding mode amplitude ratios are 
\begin{equation}
    R^\perp_{k} = \frac{h^\perp_k(f_{\rm ref})}{h^\perp_0(f_{\rm ref})}.
\label{eq:mode_ratio_perp}
\end{equation}
The normalized mode waveforms are similarly given by
\begin{equation}
    \h^\perp_k (f) = \frac{h^\perp_k(f)}{\langle h^\perp_k | h^\perp_k \rangle^{1/2}} e^{-i \arg(R^\perp_k)}.
\end{equation}

\section{Bank Model}
\label{sec:bank}
In constructing the template banks, we will first build the bank for the dominant precession harmonics and then associate each of them with four subdominant ones. Generally speaking, for each mode, we can decompose our template into the model for the amplitude and the phase
\begin{equation}
    \mathbb{h}_k(f) = A_{k}^{\rm bank}(f) e^{i \Psi_k^{\rm model} (f)} ~, \quad  k=0,1,2,3,4 ~.
\end{equation}
In the following, we will first discuss the modeling for the amplitude using \texttt{KMeans} algorithm and then we will focus on the modeling of phases using singular value decomposition (SVD) and a machine learning technique called random forest regressor (RF).

\subsection{Modeling Amplitude}
\begin{figure}[h!]
    \centering
    \includegraphics[width=1.0\linewidth]{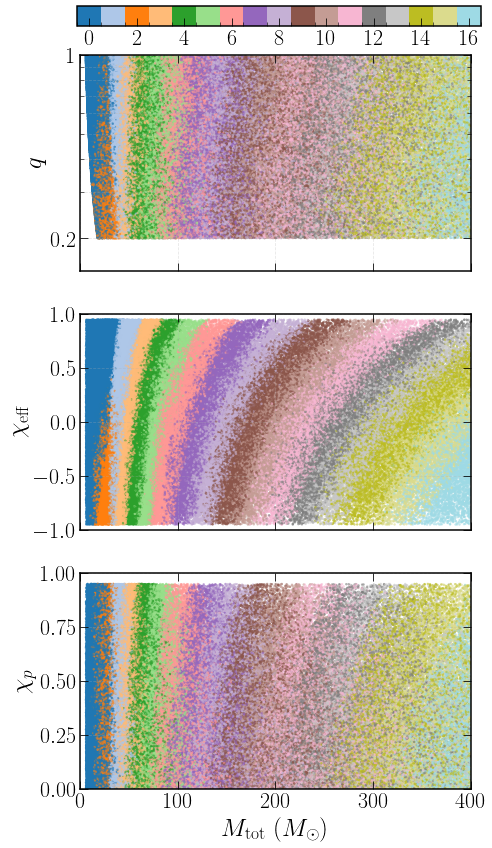}
    \caption{The template banks in our analysis are split a according to the normalized waveform amplitudes of the dominant precessing mode (the different banks are roughly distinguished by the cutoff frequencies of the waveforms). We show here the physical parameters $M_{\rm tot},q,\chi_{\rm eff},\chi_p$ correspond to different banks and the bank indices are shown with different colors.}
    \label{fig:bank_split}
\end{figure}

We would first like to divide the binary mass and spin samples into banks. We first sample the $(M_{\rm tot}=m_1+m_2, {\log q},\chi_{\rm eff},\delta \chi)$ parameter space within the range
\begin{equation}
\begin{aligned}
     3 M_{\odot} < m_2 & < m_1 < 400 M_{\odot} ~,  \\
     0.2 &< q <1 ~, \\
     |\chi_1|,|\chi_2| & < 0.99 ~,
\end{aligned}
\end{equation}
which allows us to reconstruct the physical parameter $(m_1,m_2,\chi_{1z},\chi_{2z})$. We do not extend below the range $q<0.2$ (as is used in Ref.~\cite{Dhurkunde:2026jhp}) because the higher-modes in the $L$-frame can begin to matter more than the spin precession effects, and also the \texttt{IMRPhenomXPHM} waveform fails to be accurate enough in that region due the lack of coverage from numerical relativity (NR) simulations \cite{Garcia-Quiros:2020qpx,Pratten:2020ceb,Yu:2023lml}. To incorporate the precession effect, we further sample the in-plane spin $\chi_p$ \cite{Schmidt:2014iyl}
\begin{equation}
    \chi_p \equiv \frac{1}{A_1m_1^2} {\rm max}(A_1 m_1^2 \chi_{1,\perp},A_2 m_2^2 \chi_{2,\perp}) ~,
\end{equation}
with the parameter range $[0, 0.95]$. In the above expression, $A_1 = 2 + 3 m_1 / (2m_2), A_2 = 2 + 3 m_2 / (2 m_1)$ and $\chi_{1,\perp}, \chi_{2, \perp}$ are the individual perpendicular spins. 
However, the mapping from $(\chi_{1x},\chi_{2x},\chi_{1y},\chi_{2y})$ to the single parameter $\chi_{p}$ is many to one. This means that one cannot fully reconstruct the individual spins from the $\chi_p$. As in the \texttt{IMRPhenomPv2} and \texttt{IMRPhenomXPHM} waveforms, we have made the choice of placing all of the spins on the larger black hole \cite{Schmidt:2014iyl,Pratten:2020ceb}. In our bank, we choose the convention $\chi_{1y} = \chi_p$ without loss of generality.
For each sample, we first normalize the corresponding \texttt{IMRPhenomXPHM} waveform $\langle A_0|A_0\rangle=1$. We use the \texttt{KMeans} clustering algorithm to gather all the templates with similar amplitude. The results are shown in Fig.~\ref{fig:bank_split}.
In this framework, we naturally choose the centroids of the waveform which maximizes the average overlap between amplitudes and the reference amplitude of the bank: $A_{0}^{\rm bank}$. 

\subsection{Modeling Phases}
We model the phases of waveforms based on the SVD technique \cite{Roulet:2019hzy,Wadekar:2023kym}. We first write the phases extracted from the dominant ($k=0$) mode of the \texttt{IMRPhenomXPHM} waveforms as
\begin{equation}
    \Psi_{0}^{\rm model}(f) = \langle \Psi_0\rangle_{\rm bank} + \sum_{\alpha=0}^{\rm few} c_0^\alpha \psi_\alpha^0(f) ~,
\end{equation}
where $\psi_\alpha^0$ corresponds to the orthonormal basis functions from the SVD for the dominant harmonics and $\langle \Psi_0\rangle_{\rm bank}$ denotes the average over physical waveforms in a given bank. We keep only 10 coefficients in practice. It worth noting that, in the computation of SVDs of phases, we give weights $A_0^{\rm bank}(f)^2/S_n(f)$ to the frequency bins.       
Upon using SVD, the mismatch of the phases of the dominant harmonics in the bank can be quantified as
\begin{equation}
\begin{aligned}
    & \quad \langle A_0^{\rm bank}(f)e^{i \Psi_0^{\rm model}(f;\boldsymbol{c}_0)}| A_0^{\rm bank}(f) e^{i \Psi_0^{\rm model}(f; \boldsymbol{c}_0 + \boldsymbol{\delta c}_0)}\rangle_{\rm bank} \\
    & \equiv 4 {\rm Re} \int_0^\infty df \frac{A_{0}^{\rm bank}(f)^2}{S_n(f)} e^{-i \Psi_0^{\rm model}(f;\boldsymbol{c}_0)} e^{i \Psi_0^{\rm model}(f;\boldsymbol{c}_0 + \boldsymbol{\delta c}_0)} \\
    & \simeq 1- \frac{1}{2} \sum_\alpha (\delta c_0^\alpha)^2 ~.
\end{aligned}
\end{equation}

Note that the phase difference between the high precession harmonics and the leading one is dominated by the evolution of the first Euler angle $\alpha$, i.e.
\begin{equation}
    \Psi_k^{\rm model}(f) \simeq \Psi_0^{\rm model}(f)  - k \alpha(f) ~, \quad k=1,2,3,4 ~.
\end{equation}
Therefore, instead of modeling all the harmonics independently, we just need to model the difference between the high harmonics and the dominant one
\begin{equation}
\label{eq:phase_diff}
    \Delta \Psi_k^{\rm model} \equiv \Psi_{k}^{\rm model} - \Psi_0^{\rm model} ~, \quad k=1,2,3,4.
\end{equation}
Similar to the dominant case, we perform the SVD decomposition for $\Delta \Psi_k$
\begin{equation}
   \Delta \Psi_k^{\rm model}(f) = \langle \Delta \Psi_k\rangle_{\rm bank}(f) + \sum_{\beta=0}^{\rm few} c_k^{\beta} \psi_\beta^k(f) ~.
\label{eq:SVD_decomposition}
\end{equation}
where $\psi_\beta^k$ correspond to the orthonormal basis functions for higher harmonics. Similarly, we can also quantify the mismatch of the sub-dominant precession harmonics via
\begin{equation}
\begin{aligned}
    & \quad \langle A_0^{\rm bank}(f) e^{i \Delta \Psi_k^{\rm model} (f;\boldsymbol{c}_k)}| A_0^{\rm bank}(f)e^{i \Delta \Psi_{k}^{\rm model}(f;\boldsymbol{c}_k + \boldsymbol{\delta c}_k)} \rangle \\
    & = 1 - \frac{1}{2} \sum_\beta (\delta c_k^\beta)^2 ~.
\end{aligned}
\end{equation}

\begin{figure}
    \centering
    \includegraphics[width=1.0\linewidth]{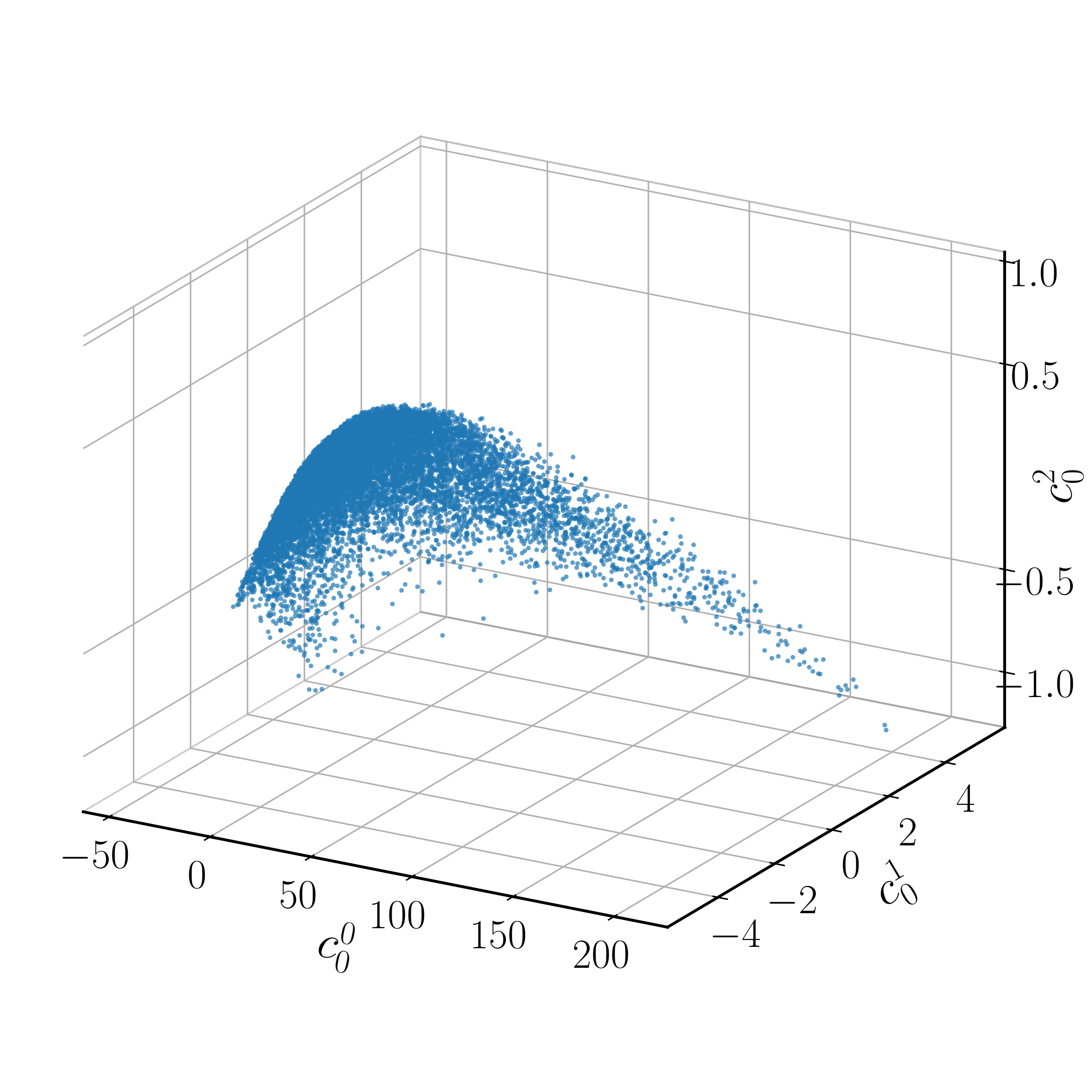}
    \includegraphics[width=0.9\linewidth]{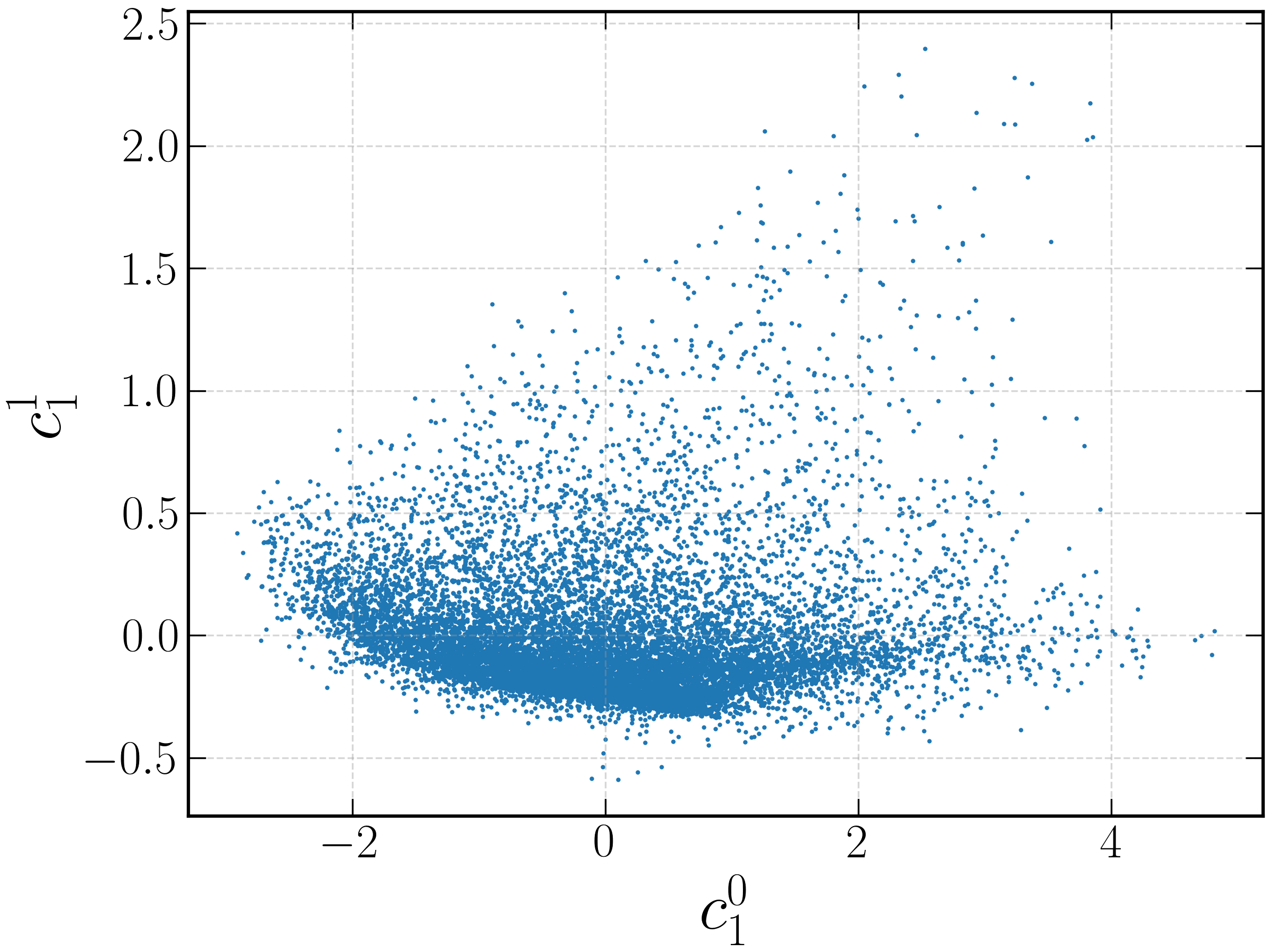}
    \caption{We model the phases of dominant-mode templates using the top three basis vectors from an SVD-based decomposition of the full GW waveforms (see Eq.~\ref{eq:SVD_decomposition}). We show here the SVD coefficients for phases in one of our banks \texttt{BBH-2}. The coefficients lie on an intrinsically lower-dimensional manifold with a non-linear structure. We therefore use a non-linear machine learning tool called random forest regressor (RF) to model the higher-order coefficients from the lowest-order and thus compress the dimensionality of our banks (see Eq.~\ref{eq:RF}).}
    \label{fig:calpha_samples}
\end{figure}

However, the $c_0^\alpha,c_k^\beta$ parameters are not fully independent. As an example, we show the distribution of $c_{0}^0,c_{0}^1,c_0^2$ and $c_1^0,c_1^1$ of the bank \texttt{BBH-2} in Fig.~\ref{fig:calpha_samples}. We see that the distribution of $c_{0}^0,c_{0}^1,c_0^2$ lives within a 2D hypersurface. Therefore, these three parameters are not independent. For higher-order precession phases, $c_1^0$ plays the dominant role when modeling $\Delta \Psi_k^{\rm model}$ in Eq.~\eqref{eq:phase_diff}. This can be understood from the semi-analytic PN expression for $\alpha(f)$ given in Eq.~\eqref{eq:alpha_f_pn}.
However, the hypersurface is not flat and does not satisfy the linear relation.
To remove the degeneracy, we use a machine learning tool called random forest regressor (RF) to reduce the parameter space \cite{Wadekar:2023kym}. We keep only three parameters $c_0^0,c_0^1,c_1^0$ and train the RF to predict the others. We train the following a separate RF corresponding to each mode:
\begin{equation}
\begin{aligned}
    \{c_0^{0},\cdots,c_0^9 \} &= {\rm RF}_0(c_0^0, c_0^1; c_1^0) ~, \\
    \quad \{c_k^0,\cdots, c_k^9\} &= {\rm RF}_k(c_0^0, \cdots c_0^9;c_1^0) ~, \quad k=1,2,3,4.
\end{aligned}
\label{eq:RF}
\end{equation}
It is worth noting that an alternative approach to dimensionality reduction is to use autoencoders \cite{Cabass:2025wqg}, but we have not yet explored it in this work.
When coming to the prediction process, we first use ${\rm RF}_0$
\begin{equation}
    \{c_0^0,c_0^1;c_1^0\} \xRightarrow{{\rm RF}_0} \{c_0^0, \cdots c_0^9\}_{\rm pred} ~,
\end{equation}
and then use ${\rm RF}_k$
\begin{equation}
    \{\{c_0^0, \cdots, c_0^9\}_{\rm pred}, c_1^0\} \xRightarrow{{\rm RF}_k} \{c_k^0, \cdots, c_k^9\} ~.
\end{equation}

To finish the construction of template banks, we need to grid the $c_0^0,c_0^1$ and $c_1^0$ space. Note that the dominant precession harmonic is generally more important than the higher precession modes, so the grid in $c_1^0$ space can be made more coarse-grained than the $c_0^0,c_0^1$ space without excessively impacting the effectualness. In Table~\ref{table:template_info}, we list the number of templates and the chirp mass range covered in each bank. In total, we have \num{57239} templates with 17 banks, roughly two times larger than a typical aligned-spin BBH $(2,2)$-only bank, see e.g., \cite{Wadekar:2023kym}. But, it is much smaller than the template banks built from directly sampling the physical parameters in Ref.~\cite{Schmidt:2024jbp}.

\begin{center}
\begin{table}[t]  
\centering
\setlength{\tabcolsep}{6pt}
\renewcommand{\arraystretch}{1.2}
\rowcolors{2}{gray!12}{white}
\begin{tabular}{l p{0.2\linewidth} p{0.4\linewidth}}
\hline\hline
Bank & $N_{\mathrm{templates}}$ & $\mathcal{M}_{\rm chirp}$ range (M$_\odot$) \\
\hline
BBH-0  & \num{37845} & [2.6,\,15.9] \\
BBH-1  & \num{2061}  & [9.1,\,25.2] \\
BBH-2  & \num{12110} & [3.9, 13.2] \\
BBH-3  & \num{1825}   & [11, 31.6] \\
BBH-4  & \num{1214}   & [12.3, 38.4] \\
BBH-5  & \num{722}   & [14.5, 47.2] \\
BBH-6  & \num{583}   & [18.9, 57.6] \\
BBH-7  & \num{434}   & [22.7, 69.6] \\
BBH-8  & \num{216}    & [28.3, 86.7] \\
BBH-9  & \num{47}    & [34.8, 101.8] \\
BBH-10 & \num{47}    & [40.3, 121.6] \\
BBH-11 & \num{31}    & [47.4, 141.7] \\
BBH-12 & \num{26}    & [54.6, 164.6] \\
BBH-13 & \num{18}    & [62.7, 172.7] \\
BBH-14 & \num{21}    & [71.7, 173.7] \\
BBH-15 & \num{18}    & [82.3, 174.1] \\
BBH-16 & \num{21}    & [93.7, 173.8] \\
\hline
Total  & \num{57239} & \\
\hline\hline
\end{tabular}
\caption{Details of our template bank for the dominant precession harmonics. We split the bank according to the amplitude of the dominant precession harmonics using \texttt{KMeans} algorithm. In the second column, we show the number of templates in each bank. In the third column, we list the range of the chirp mass in each of the bank.}
\label{table:template_info}
\end{table}
\end{center}

\section{Matched filtering, marginalization and normalizing flow}
\label{sec:marginalize}

\subsection{Mode-by-mode matched filtering}

In this section, we calculate the effectualness of our precession banks. We first perform matched-filtering of each harmonic-mode template separately with the data $d(f)$ and collect the complex mode-by-mode SNR timeseries:
\begin{equation}
\label{eq:mode_SNR}
    \rho_{k}(t) = \langle \mathbb{h}_k(f)| d(f) e^{i 2\pi ft} \rangle ~,
\end{equation}
where $\mathbb{h}_k(f)$ is the normalized template waveform for mode $k$. From the waveform model constructed in Eq.~\eqref{eq:pres_harm}, we expect that the SNR ratio scales as 
\begin{equation}
    |\rho_k|/|\rho_0| \sim b^k < 1 ~, \quad k=1,2,3,4~,
\end{equation}
This behavior is encoded in our mode-ratio prior samples $R_k$. We therefore want to construct a detection statistic which combines the different mode SNRs along with the mode ratio priors. We use the Neyman--Pearson lemma \cite{neyman1933ix}, which states that the optimal detection statistic is the likelihood ratio for the entire
data associated with the trigger under the signal $\mathcal{S}$ and noise $\mathcal{N}$ hypotheses:

\begin{equation}
    \frac{P(d|\mathcal{S})}{P(d|\mathcal{N})} =    \frac{P(d|\mathcal{S})}{P(d|\mathrm{GN})} \frac{P(d|\mathrm{GN})}{P(d|\mathcal{N})}
\end{equation}
where we have deliberately broken the statistic into two parts and denote the hypothesis when the noise is Gaussian by GN. The non-Gaussian correction to the detection statistic is given by the second term. We refer the reader to Ref.~\cite{Wadekar:2024zdq} for more details on the calculation of the non-Gaussian correction (see also \cite{Venumadhav:2019tad,psd_drift, Venumadhav:2019lyq}). To calculate the first term, we marginalize over the free parameters inherent to the signal hypothesis.
\begin{equation}
\begin{aligned}
\label{eq:marg_integral}
    & \quad \frac{P(d|\mathcal{S})}{P(d|\mathrm{GN})} \\
    & = \frac{\int \dd \Pi(\theta_i,\theta_e) P(\theta_i,\theta_e) \exp\{-\langle d-h)|d-h\rangle/2\}}{\exp\{-\langle d|d\rangle/2\}}  \\
    & = \int \dd \Pi(\theta_i,\theta_e) P(\theta_i,\theta_e) \exp\Big[ {\rm Re}(\langle d| h \rangle) - \frac{1}{2} \langle h | h \rangle \Big] ,
\end{aligned}
\end{equation}
where $h$ here is the waveform model of intrinsic $\theta_i \in \{m_1, m_2, \vec{\chi}_1, \vec{\chi}_2\}$ and extrinsic parameters $\theta_e \in \{t_d, \theta_\mathrm{JN}, \phi_0, D, \psi, \hat{\bm{n}}\}$, which correspond to detector arrival time, inclination, initial orbital phase, luminosity distance, polarization and sky position (right ascension and declination) respectively $P(\theta_i,\theta_e)$ and $ \Pi(\theta_i,\theta_e)$ are the prior density and phase space differential volume respectively. Ideally, to perform the marginalization over intrinsic parameters, one can do a Monte Carlo sum over the contribution from all the templates in the bank. However, this is computationally expensive. Instead, we can use the best-fit template to approximate the marginalization. An approximate version of the marginalization statistic can be computed as
\begin{equation}
\begin{aligned}
\label{eq:margin_like}
&\frac{P(d|\mathcal{S})}{P(d|\mathrm{GN})} \simeq P_{\theta_{i,\mathrm{max}}}\int \dd \Pi(R_1,R_2,R_3,R_4|\theta_{i,\mathrm{max}})\Pi(\theta_e) \\
    & \quad \times \exp \left[ \sum_{k\in \mathrm{detectors}} \mathrm{Re}(\langle h_k | d_k \rangle) - \frac{1}{2}\langle h_k | h_k \rangle \right] ~,
\end{aligned}
\end{equation}
up to logarithmic correction, where $P_{\theta_{i,\mathrm{max}}}$ is the astrophysical prior corresponding to
the best-fit template.

\subsection{Coherent score in Gaussian noise}
\label{sec:CoherentScore}
The marginalization integral in Eq.~\eqref{eq:margin_like} is calculated using the marginalization algorithm implemented in the publicly available \texttt{cogwheel} package \cite{Rou22_cogwheel,Rou23_CoherentScore,cogwheel}.
Using the formula in Eq.~\eqref{eq:harm_wave} for a particular template from the template bank, the predicted waveform $h_k$ of the signal in detector $d$ becomes
\be \begin{aligned}\label{eq:predictedwf}
  h_d (f) & \simeq \frac{D_0}{D} e^{2 i \phi_0} \sum_k \left[ F_{+, d} \A_{d,k}^+  -  i F_{\times, d} \,\A_{d,k}^\times \right] \\
  & \quad \times R_{k}\ \mathbb{h}_{k} (f)~,
\end{aligned}
\ee
where $\mathbb{h}$ corresponds to complexified unit templates ($\rm SNR=1$) for different modes and $D_0$ is the distance where $\rm SNR=1$ for the dominant mode (i.e. $k=0$). $F_{p, k}$ is the antenna response of the detector to polarization $p \in \{+,\times\}$ and is dependent on the polarization angle $(\psi)$ and sky location $(\hat{\bm{n}})$.  $R_{k}$ are the mode ratio samples corresponding to variation of the ratio of the SNR in the $k$-th mode to that in the dominant mode over intrinsic parameters in the subbank, see Eq.~\eqref{eq:mode_ratio}. 

Let us now calculate the inner products $\langle h|d\rangle$ and $\langle h|h\rangle$ used in Eq.~\eqref{eq:margin_like} for a particular detector $d$:
\begin{widetext}
\be \begin{split}
  \langle h (f) | h (f) \rangle =& \sum_{k, k'} \langle
  h_{k}(f) | h_{k'}(f) \rangle\\ =& \sum_{k, k'}  \frac{D^2_0}{D^2}\left[ F_{+, d} \A_{d,k}^+ \, + \, i F_{\times, d} \,\A_{d,k}^\times \right] \times \left[ F_{+, d} \A_{d,k'}^+ \, - \, i F_{\times, d} \,\A_{d,k'}^\times \right]\,  R^*_{k} R_{k'} \langle\h_{k}(f)|\h_{k'}(f)\rangle
\end{split}\label{eq:hh}\ee
and
\be \begin{split}
\langle h (f) | d_d (f)& e^{i2\pi f t}  \rangle = \sum_{k} \langle
  h_{k} (f) | d_d (f) e^{i2\pi f t} \rangle\\
  =& \frac{D_0}{D} \sum_{k} \left[ F_{+, d} \A_{d,k'}^+ \, + \, i F_{\times, d} \,\A_{d,k'}^\times \right] e^{-2i\phi_0} R^*_{k}\, \rho_{k,d}(t),
\end{split} \label{eq:dh} \ee
where we have denoted the complex inner product of the data with the unit template in detector $k$ as $\rho_{k,d}(t)$. The expected time of arrival of the signal in each detector is given by $t_k \equiv t_\oplus + \bm{r}_k\cdot \hat{\bm{n}}/c$, which is the arrival time at the geocenter plus a correction depending on the location of detector $\bm{r}_d$. We compute the integral in Eq.~\eqref{eq:margin_like} for given a template as
\be
\frac{P(d|\mathcal{S})}{P(d|\mathrm{GN})} \simeq P_{\theta_{i,\mathrm{max}}} \sum_{i}w^{(i)}
 \int d\Pi(\theta_\mathrm{JN}, \phi_0, D, \psi, \hat{\bm{n}}) \exp \left[ \sum_{d\in \mathrm{detectors}} \ln \mathcal{L}_d (t_d, R^{(i)}_{k}, \theta_\mathrm{JN}, \phi_0, D, \psi, \hat{\bm{n}}) \right]
\label{eq:MultiDetCS}
\ee
\end{widetext}
where the log-likelihood is $\ln \mathcal{L}_d \equiv \mathrm{Re}(\langle h_d | d_d \rangle) - \frac{1}{2}\langle h_d | h_d \rangle$, and we calculate the marginalization over $D, \phi_0$, $\hat{\bm{n}}, \psi, t_\oplus$ and $R_{k}$. Let us first focus on calculating the integral over mode amplitude ratios. In Eq.~\eqref{eq:MultiDetCS} we use Monte Carlo integration using the $R_{k}$ samples corresponding to a particular template generated using a conditional normalizing flow, as discussed in the next section and shown in Fig.~\ref{fig:nflow}.
$w$ is again the weight of each sample corresponding to its observable volume (e.g., samples with asymmetric mass ratios have lower observable volumes and vice versa).

Algorithms to calculate each of the remaining integrals are discussed in detail in \citet{Rou23_CoherentScore}. We integrate over $D$ by interpolating a precomputed table, over $\phi_0$ analytically, and over the remaining extrinsic parameters $\theta_\mathrm{JN}, \hat{\bm{n}}, \psi$ using adaptive importance sampling. Overall, given a multi-detector trigger, our pipeline currently needs $\sim 0.2$ s for calculating the full marginalization integral, and thus is computationally-feasible for a search algorithm using coincident and time-slided GW strain data.

\subsection{Normalizing flow for template-dependent mode ratio samples}

\begin{figure}
    \centering
    \includegraphics[width=1.0\linewidth]{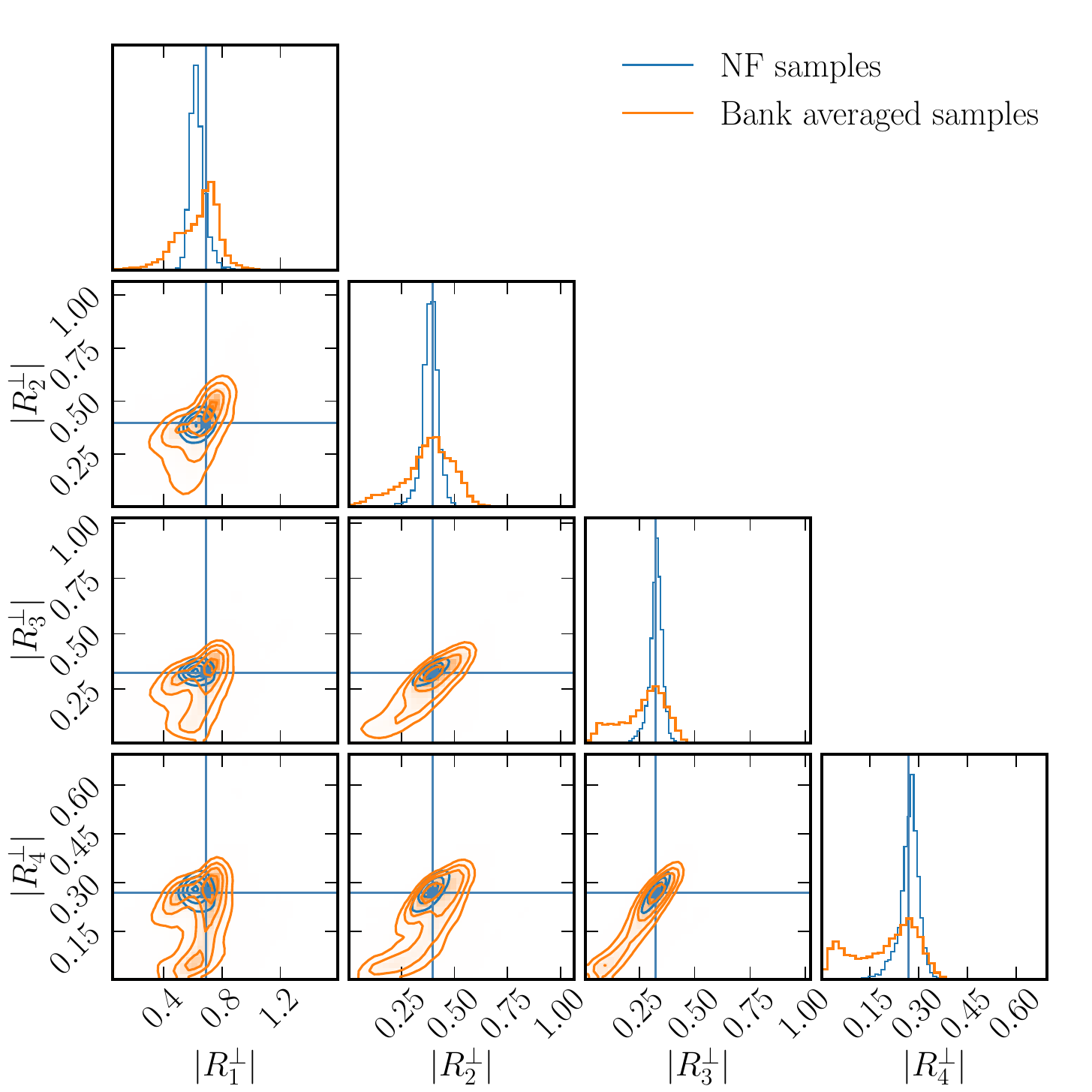}
    \caption{Each of the dominant-mode template in our banks is accompanied by $(i)$ normalized sub-dominant mode templates and $(ii)$ mode-SNR-ratio samples corresponding to the chunk of parameter phase space associated with the dominant template [$R^\perp_{k} = h^\perp_k(f_{\rm ref})/h^\perp_0(f_{\rm ref})$ for precession harmonic $k$, see Eq.~\eqref{eq:mode_ratio_perp}]. We model the distribution of the mode-ratio samples $\boldsymbol{R}$ using a conditional normalizing flow (NF). We show the mode ratio amplitude distribution $p_\theta(|\boldsymbol{R}|\, \vert\, \boldsymbol{c})$ (in blue) learned from NF. The blue line denotes the point true value of $\boldsymbol{R}$ for a particular choice of intrinsic parameters in one of our banks \texttt{BBH-7}. Compared with the $|\boldsymbol{R}|$ distribution corresponding to the entire bank, NF significantly shrinks the prior domain for the marginalization integral over $|\boldsymbol{R}|$, which improves the efficiency of marginalization.}
    \label{fig:nflow}
\end{figure}

Each of our templates corresponds to a particular region in the parameter space of \{$m_1, m_2, \vec{\chi}_1,\vec{\chi}_2 $\}. While each point in this region gives roughly similar normalized modes $\h_k$ waveforms, the ratios of the modes $R_k$ from Eq.~\ref{eq:mode_ratio} do vary.
We use normalizing flows to model the distribution of $R_k$ corresponding to each template.
Essentially, we want to learn the conditional distribution
\begin{equation}
    p_\theta (\boldsymbol{R}|\boldsymbol{c}) ~,
\end{equation}
where $\boldsymbol{R} = (R_1^\perp,R_2^\perp,R_3^\perp,R_4^\perp)$ and $\boldsymbol{c} = (c_0^0, c_0^1, c_1^0)$ is the set of coefficients corresponding to a template in the bank. In the conditional NF, this is achieved by defining an invertible map to a standard normal distribution
\begin{equation}
    z = f_\theta(\boldsymbol{R};\boldsymbol{c}) ~, \quad z \sim \mathcal{N}(0,1) ~.
\end{equation}
Through the changing of variable, this map generates the following relation
\begin{equation}
\log p_\theta(\boldsymbol{R}\,|\,\boldsymbol{c})
=\log p(z)
+\log\left|\det\frac{\partial f_\theta(\boldsymbol{R};\boldsymbol{c})}{\partial \boldsymbol{R}}\right|.
\end{equation}
We train $f_\theta$ by maximizing the conditional log-likelihood. This procedure is better than directly modeling the distribution $p_\theta$ using neural networks because it guarantees non-negativity, normalizability, and it is much easier to sample. Since $\boldsymbol{R}$ is complex in our scenario, we model the amplitude and phase distributions jointly. In Fig.~\ref{fig:nflow}, we plot the learned amplitude distribution corresponding to one sample in bank \texttt{BBH-7}. The true value of $\boldsymbol{R}$ is shown as a blue dot in the middle of the distribution. Compared with the $|\boldsymbol{R}|$ distribution combining all the samples from the entire template bank, we see that the NF increases the marginalization efficiency by fine-tuning the distribution of the $|\boldsymbol{R}|$ samples for each template in the bank.

\subsection{Approximate single-detector marginalized statistic}

In a typical matched-filtering based search pipeline, before we do a coincidence search from multi-detector data, we first need to collect triggers by filtering the templates with strain from individual detectors and then compare the triggers from different detectors to form a coincidence trigger. However, in the first stage of the pipeline, we do not have information about the triggers in the other detectors, and we cannot directly use the multi-detector coherent statistic in Eq.~\eqref{eq:MultiDetCS}. We therefore generalize the results for the single detector case in this section.

We need the single-detector statistic to be quick to
compute as it will be applied to a large number of low
SNR noise triggers.
We therefore combine all the extrinsic marginalized integrals into a single Monte Carlo sum over re-defined prior samples given by 
\be
R^F_k \equiv \frac{[F_{+} \mathcal{A}_k^{+}-i F_{\times} \mathcal{A}_k^{\times}]\, R^\perp_k}
{[F_{+} \mathcal{A}_0^{+}-i F_{\times} \mathcal{A}_0^{\times}]\, R^\perp_0}
\ee
where we have now included the detector response terms in the mode ratio samples.
The full waveform can thus be written in a simple manner as
\be
h(f) = \mathcal{C} \sum_{k=0}^4\, R^F_k\, \h^\perp_k(f)
\label{eq:waveform_response_ratio_decomposition}
\ee
where $\mathcal{C} = [F_{+} \mathcal{A}_0^{+}-i F_{\times} \mathcal{A}_0^{\times}] e^{i2 \phi_0} D_0/D$.
Practically, we perform the marginalization over complex ratio $R^F_k$ by using Monte Carlo sampling. To get the prior samples of $R^F_k$, we simulate the \texttt{IMRPhenomXPHM} waveforms with the following inclination distribution \cite{Schutz:2011tw}
\begin{equation}
P(\theta_\mathrm{JN})=\left[\frac{1}{8}\left(1+6 \cos ^2 \theta_\mathrm{JN}+\cos ^4 \theta_\mathrm{JN}\right)\right]^{3 / 2} \sin \theta_\mathrm{JN}
\end{equation}
and randomly sampled sky location $\Theta,\Phi$ and polarization $\psi$. The GW strain at the detector can be written as 
\begin{equation}
    \begin{split}
        h^{\rm det}(f) =\ &h_+^{\rm XPHM}(f) F_+(\Theta,\Phi,\psi) \\
        & + h_\times^{\rm XPHM}(f) F_\times(\Theta,\Phi,\psi) ~.
    \end{split}
\end{equation}
Then the ratio $R_k$ can be computed by taking the inner product of the full waveform with the orthogonalized mode templates (i.e. $\mathbb{h}^\perp_k(f)$) as
\begin{equation}
    R^F_{k} = \frac{\langle \mathbb{h}^\perp_k(f)| h^{\rm det}(f) \rangle}{\langle \mathbb{h}^\perp_0 (f)| h^{\rm det}(f) \rangle} ~.
\label{eq:effective_mode_ratios}
\end{equation}

\begin{figure*}[htb]
    \centering
    \includegraphics[width=0.48\textwidth]{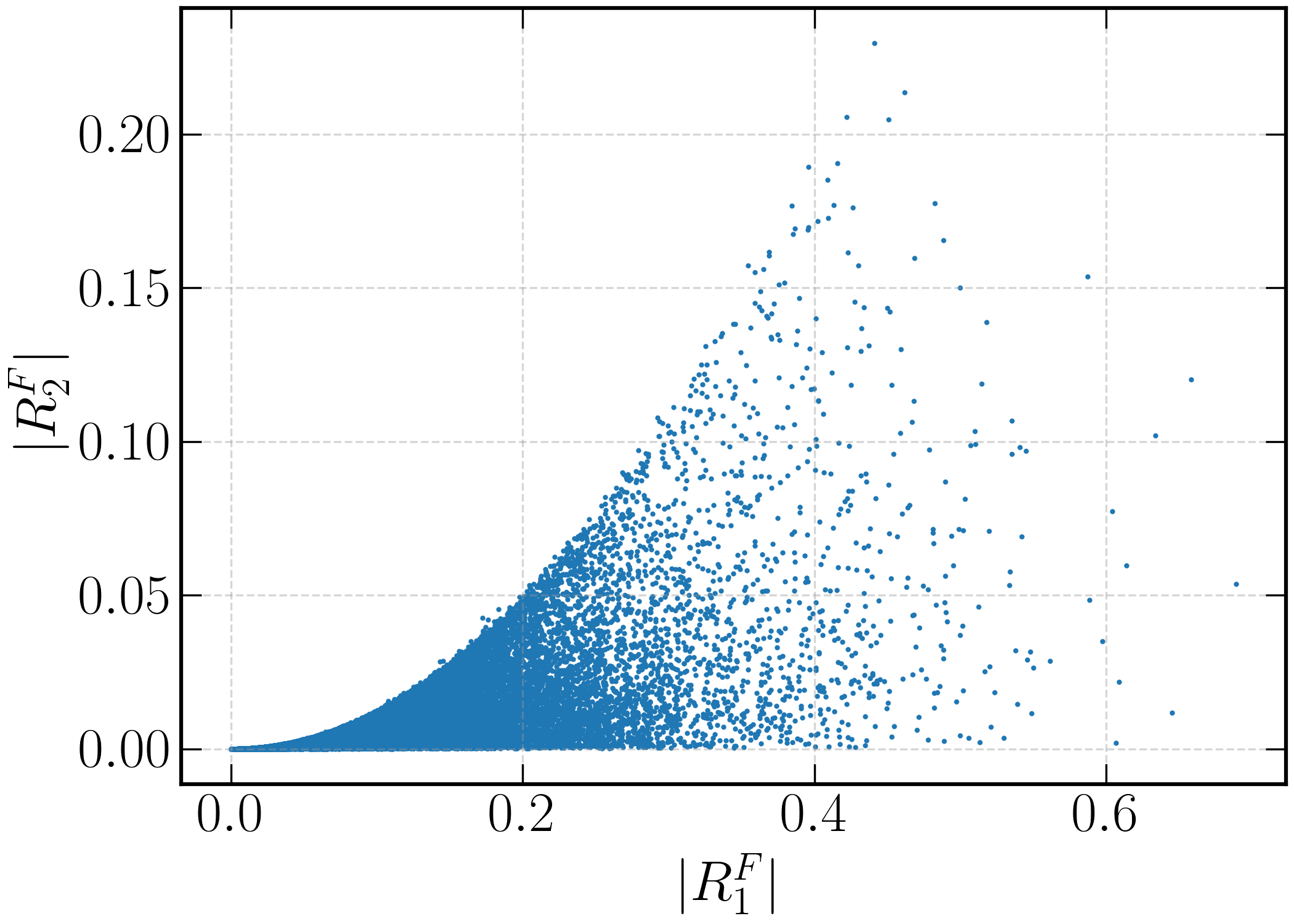}\hfill
    \includegraphics[width=0.48\textwidth]{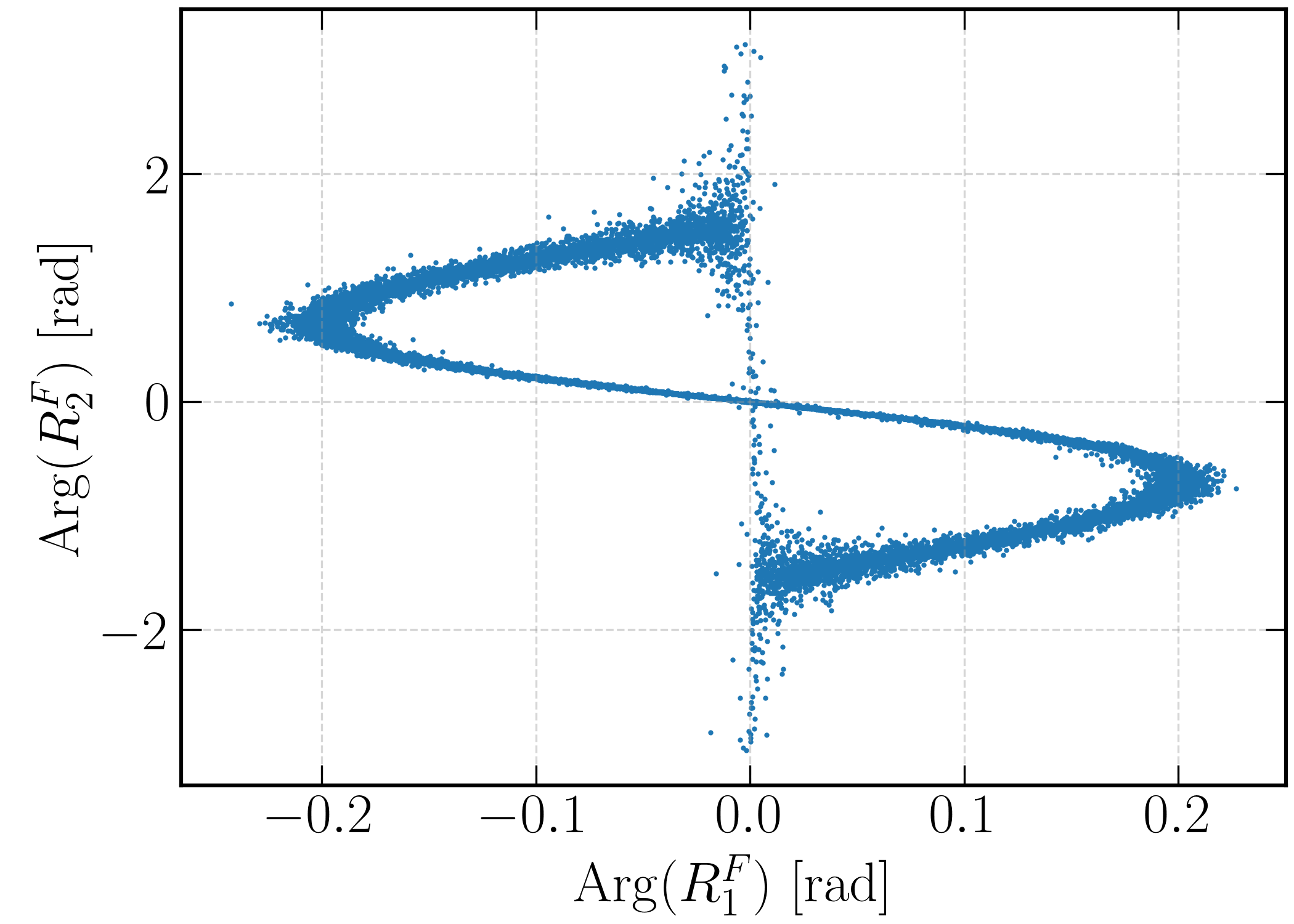}
    \caption{Samples of the effective precession mode ratios (including detector response, see Eq.~\ref{eq:effective_mode_ratios}) $R^F_{1},R^F_{2}$ of the bank \texttt{BBH-0}. \textbf{Left}: The amplitude of the mode ratio. $|R^F_2|$ is generally smaller than $|R^F_1|$ because of the suppression from precession opening angle, i.e. $b(f)<1$ in Eq.~\eqref{eq:pres_harm}. \textbf{Right}: The phase of the mode ratio. The main contribution for the variation comes from the $\alpha(f_{\rm ref})$ and the inclination angle dependence in Eq.~\eqref{eq:A_angle}.}
    \label{fig:placeholder}
\end{figure*}

We show the distribution of mode ratio amplitudes $|R^F_1|, |R^F_2|$ and the corresponding phases $\arg(R^F_1), \arg(R^F_2)$ of bank \texttt{BBH-0} in Fig.~\ref{fig:placeholder}. 
In the left panel, we see that the samples mostly fall within the region $|R^F_1|>|R^F_2|$, which aligns with the hierarchy of the precession harmonics dictated by $b<1$ given in Eq.~\eqref{eq:pres_harm}. The right panel reflects the distribution of inclination angles given by Eq.~\eqref{eq:A_angle}.

With the samples of the mode ratio coefficients $R_k^*$, we can explicitly evaluate the marginalized likelihood in Eq.~\eqref{eq:margin_like}. Plugging our waveform model in Eq.~\eqref{eq:waveform_response_ratio_decomposition} into Eq.~\eqref{eq:marg_integral}, we can get
\begin{widetext}
\begin{equation}
\label{eq:marg_deri_1}
    \exp\Big(\frac{\rho_{\rm marg}^2}{2}\Big)  = \int d\Pi(R^F_{1},R^F_2,R^F_3,R^F_4,D) \exp \Bigg(N_{\rm det} \Bigg[{\rm Re} \Bigg\{\mathcal{C}^* (\rho_{0}^\perp + \sum_{i=1}^4 R_i^{F\,*} \rho_i^\perp)\Bigg\} - \frac{1}{2} |\mathcal{C}|^2 (1 + \sum_{i=1}^4 |R^F_i|)^2\Bigg]\Bigg) ~.
\end{equation}    
where $\mathcal{C} = [F_{+} \mathcal{A}_0^{+}-i F_{\times} \mathcal{A}_0^{\times}] e^{i2 \phi_0} D_0/D$ is nuisance complex parameter which we will maximize over in our single-detector statistic. The value of $\mathcal{C}_{\rm max \mathcal L}$ where the likelihood gets maximized is
\begin{equation}
    \mathcal{C}_{\rm max \mathcal L} = \frac{\rho_0^\perp + \sum_{i=1}^4 R^F_i \rho_i^\perp} {1 + \sum_{i=1}^4 |R^F_i|^2} ~.
\end{equation}
Plugging this equation back into Eq.~\eqref{eq:marg_deri_1}, we arrive at
\begin{equation}
    \exp\Big(\frac{\rho_{\rm marg}^2}{2}\Big) = \int d\Pi(R^F_1,R^F_2,R^F_3,R^F_4) \exp\Bigg[\frac{N_{\rm det}}{2(1 + \sum_{i=1}^4 |R^F_i|^2)} \Big|\rho_{0}^\perp + \sum_{i=1}^4 R_i^{F\,*} \rho_i\Big|^2\Bigg]~.
\end{equation}
This integral can be performed via Monte Carlo sampling with the following equation
\begin{equation}
\label{eq:margin_single_det}
\exp \Big(\frac{\rho_{\rm marg}^2}{2}\Big) \simeq P_{\rm prior} \sum_{j \in {\rm samples}} w^{(j)} \exp\Bigg[\frac{N_{\rm det}}{2(1 + \sum_{i=1}^4 |R_i^{F\, (j)}|^2)} \Big|\rho_{0}^\perp + \sum_{i=1}^4 R_i^{F*\, (j)} \rho_i\Big|^2\Bigg] ~,
\end{equation}
\end{widetext}
where we sum all the samples for the mode ratio with weights $w^{(j)}$ corresponding to the observable volume of the sample (e.g., samples with nearly edge-on inclinations or low $q$ have lower volume and vice versa). 

\subsection{Comparing marginalization and maximization with mock data}

\begin{figure}
    \centering
    \includegraphics[width=1.0\linewidth]{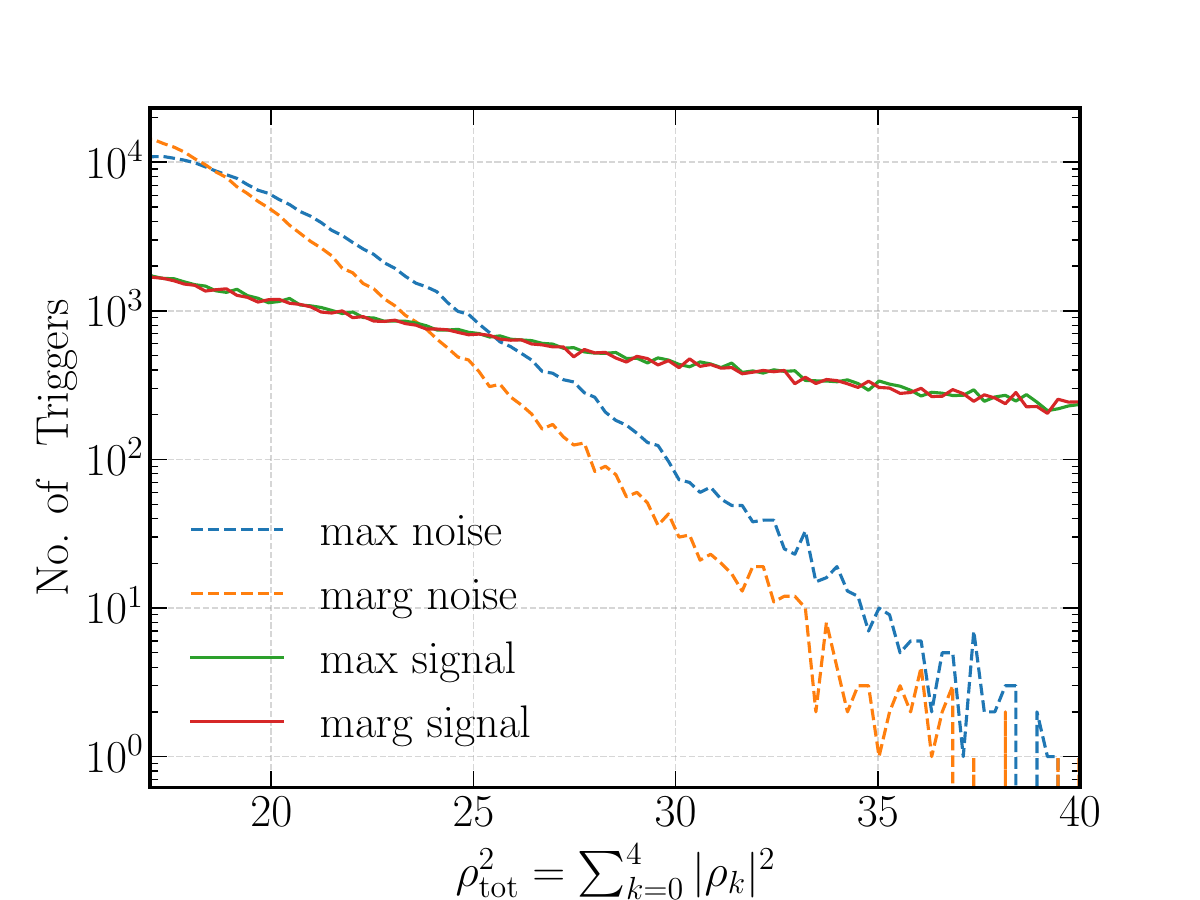}
    \includegraphics[width=1.0\linewidth]{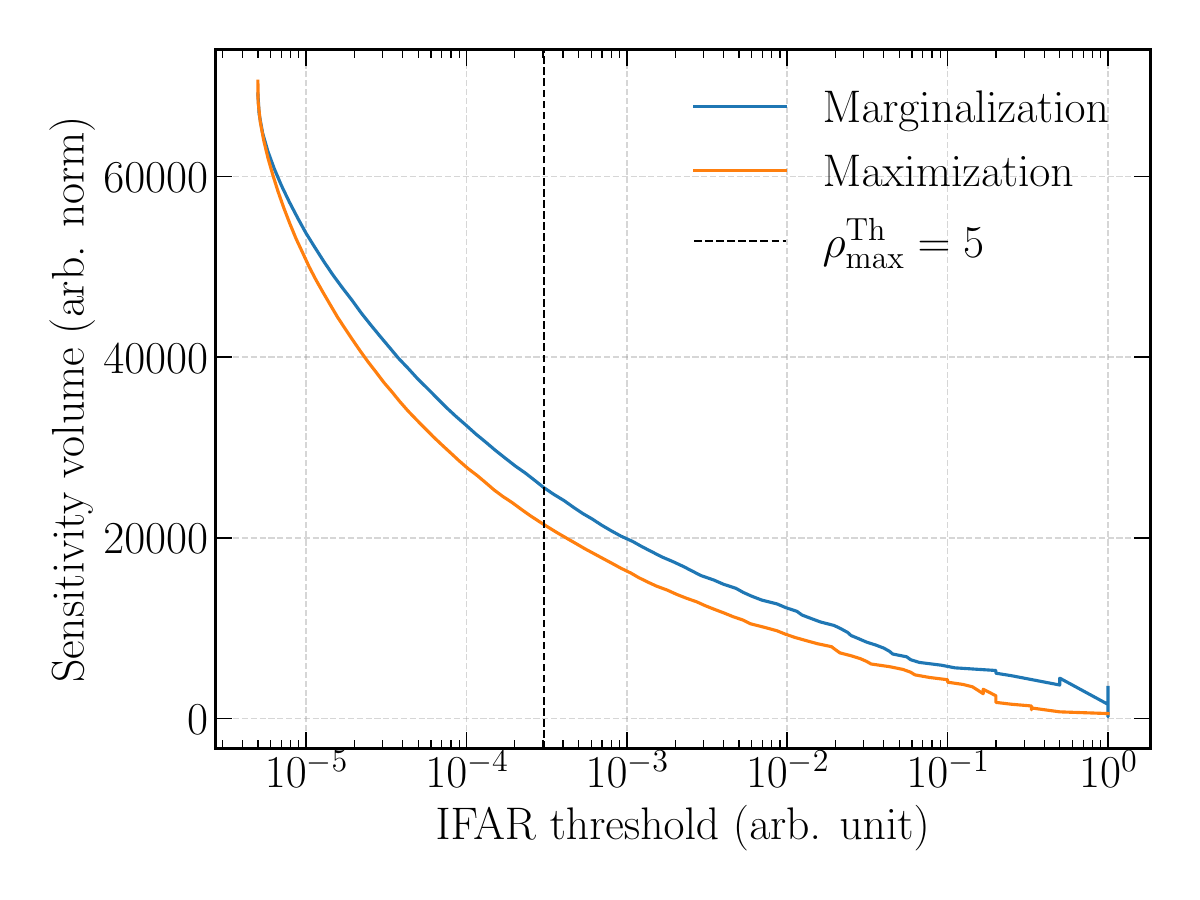}
    \caption{Mock detection statistics sampled from noise trigger distribution in Eq.~\eqref{eq:noise_dist} and Eq.~\eqref{eq:signal_dist}. \textbf{Top}: The marginalization and maximization statistics are calculated according to the noise distribution Eq.~\eqref{eq:margin_single_det} and the signal distribution Eq.~\eqref{eq:max_single_det} respectively. \textbf{Bottom}: The number of recovered candidates, i.e. the sensitivity volume given the inverse false alarm rate (IFAR) threshold computed from the upper panel. The marginalized values are generally higher than the maximized values. For instance, the gain in the sensitive volume for the SNR threshold $\rho_{\rm max}^{\rm Th} = 5$ is $\simeq 8.8\%$.}
    \label{fig:mock_stat}
\end{figure}

In this section, we show that the marginalizing over the mode-amplitude ratios in Eq.~\eqref{eq:margin_single_det} improves the sensitive volume compared to maximimizing over the mode-ratios (as used in previous search algorithms, e.g., \cite{Dhurkunde:2026jhp}). We use the following statistic for the maximization case:
\begin{widetext}
\begin{equation}
\label{eq:max_single_det}
    \exp \Big(\frac{\rho_{\rm max}^2}{2}\Big) \simeq P_{\rm prior} \max_{\rm j \in {\rm samples}} \exp\Bigg[\frac{N_{\rm det}}{2(1 + \sum_{i=1}^4 |R_i^{F\, (j)}|^2)} \Big|\rho_{0}^\perp + \sum_{i=1}^4 R_i^{F\,*\,(j)} \rho_i\Big|^2\Bigg] 
\end{equation}
\end{widetext}
To quantify the difference between these two methods, we study the following simple mock detection statistics. With Gaussian noise, the complex SNR distribution for the noise triggers follows the following distribution
\begin{equation}
\label{eq:noise_dist}
    P_{\rm noise}(\rho_k) \propto e^{-|\rho_k|^2/2} ~, \quad k=0,1,2,3,4~.
\end{equation}
For the signals, the distribution becomes \cite{Schutz:2011tw}
\begin{equation}
\label{eq:signal_dist}
    P_{\rm signal} (\rho_k) \propto |\rho_k|^{-4} ~.
\end{equation}
We give the derivation of this distribution in Appendix~\ref{app:signal_dist}.
After sampling these distributions and marginalizing (maximizing) over the mode ratios using Eq.~\eqref{eq:margin_single_det} (Eq.~\eqref{eq:max_single_det}), we show the detection statistics in Fig.~\ref{fig:mock_stat}.
For a fixed number of noise triggers, we can get different thresholds for the marginalized and maximized cases. In other words, we fix
\begin{equation}
\label{eq:Noise_fix}
    N_{\rm noise}(\rho_{\rm marg}> \rho_{\rm marg}^{\rm Th}) = N_{\rm noise} (\rho_{\rm max} > \rho_{\rm max}^{\rm Th})  ~,
\end{equation}
which naturally indicates 
\begin{equation}
    \rho_{\rm marg}^{\rm Th} < \rho_{\rm max}^{\rm Th} ~.
\end{equation}
As a benchmark, we choose $\rho_{\rm max}^{\rm Th} = 5$ in the analysis. Within the condition Eq.~\eqref{eq:Noise_fix}, we calculate the ratio of the signal triggers and find
\begin{equation}
    \frac{N_{\rm signal}^{\rm marg}}{N_{\rm signal}^{\rm max}} -1 \simeq 8.8\% ~,
\end{equation}
which indicates that we get $8.8\%$ increase in the sensitivity volume at a fixed false alarm threshold when marginalizing over the mode-amplitude ratios. We show the results for different thresholds in Fig.~\ref{fig:mock_stat} and roughly find a similar level of improvement. Note that we have not marginalized over distance and reference orbital phase in the simple toy example in this section. Our coherent multi-detector statistic in section~\ref{sec:CoherentScore} does include marginalization over all extrinsic parameters, and hence we expect the sensitivity improvement to be even larger for that case.

\section{Calculating Effectualness of Banks}
\label{sec:effectualness}
\begin{figure}
    \centering
    \includegraphics[width=1.0\linewidth]{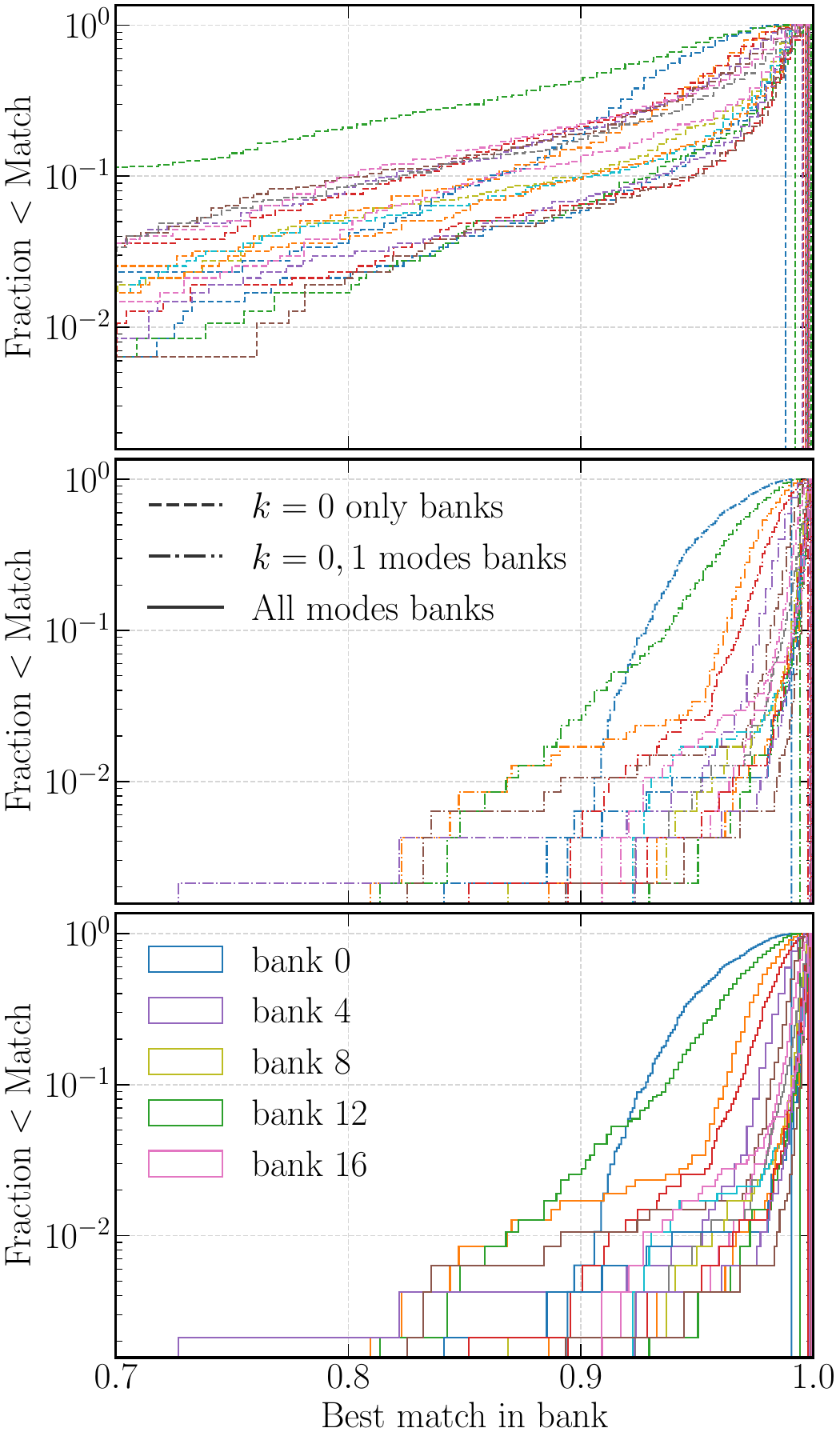}
    \caption{Effectualness of our template bank tested on random waveforms drawn from \texttt{IMRPhenomXPHM} with $(\ell=2,m'=2)$ mode in the L-frame. The y-axis shows the fraction of the test samples with match less than the value on the x-axis. \textbf{Top}: We show the match using the $k=0$ only banks.
    \textbf{Middle}: The effectualness with two precession harmonics included. There is a significant improvement upon including the $k=1$ sub-dominant harmonic.
    \textbf{Bottom}: Results for effectualness with all precession harmonics included. As seen in Fig.~\ref{fig:volume_loss_precession_harmonics}, including three or more harmonics in our templates does not further improve the match for our search (which is restricted to $q>0.2$).}
    \label{fig:effectualness}
\end{figure}

In this section, we are going to quantify the effectualness of our precession template banks. Based on Eq.~\eqref{eq:mode_SNR}, we replace the data $d$ with normalized test waveform from \texttt{IMRPhenomXPHM} with $(2,2)$ mode in $L$-frame. For future convenience, we define the normalized SNR of each mode $k$ to be
\begin{equation}
    \tilde \rho_k^\perp (t) = \frac{1}{\sqrt{\langle h^{\rm XPHM}|h^{\rm XPHM} \rangle }}\langle \mathbb{h}_k^\perp (f)| h^{\rm XPHM} (f) e^{i 2\pi f t}\rangle ~.
\end{equation}
The match of the whole waveform can be quantified as the maximized summation of SNRs in all precession harmonics, i.e.
\begin{equation}
\begin{aligned}
    {\rm match}^2 & = \max _t \Big(|\tilde \rho_0^\perp(t)|^2 + |\tilde \rho_1^\perp(t)|^2 + |\tilde \rho_2^\perp(t)|^2 \\
    & \quad + |\tilde \rho_3^\perp(t)|^2 + |\tilde \rho_4^\perp(t)|^2\Big) ~.
\end{aligned}
\end{equation}
We show the effectualness of our template bank in Fig.~\ref{fig:effectualness}.
Overall, we obtain matches $\geq 90\%$ for $99\%$ of the test waveforms. To save the computational cost, instead of computing the inner product with each template in the bank, we first project the noiseless test waveform into $\{c_0^0,c_0^1;c_1^0\}$ space by projecting with the SVD bases. Then, we find the closest grid point in the $\{c_0^0,c_0^1;c_1^0\}$ and use RFs to reconstruct the precession harmonics and then calculate the match. The match of \texttt{BBH}-0 and \texttt{BBH}-2 is sightly worse than the other banks. This is because we use relative large parameter spacing $\Delta c=0.7$ for \texttt{BBH}-0 and $\Delta c= 0.5$ for \texttt{BBH}-2 to reduce the number of templates. The mismatch will be smaller if we shrink $\Delta c$. We show more details in Appendix \ref{app:bank_spacing}.

\section{Conclusion and Future Directions}
\label{sec:con_future}

In this work, we have constructed a new template bank to search for precessing binaries based on precession harmonic decomposition for gravitational wave signals. For each harmonic, we have used the \texttt{KMeans} algorithm, SVD decomposition, and RF machine learning techniques. By filtering the precession harmonics separately and then combining their contributions to construct the marginalized likelihood, we are able to combine information across various modes without substantially enlarging the template bank. This greatly reduces the computational cost compared to brute-force sampling over sky locations and inclination angles. The matched filtering stage could be further accelerated by precomputing the inner products $\langle d | h_\alpha \rangle$ against the SVD basis vectors once, and then obtaining the SNR timeseries for all templates via a single matrix multiplication over the basis coefficients $c_\alpha$ following the spirit of the \texttt{dot-PE} framework~\cite{Mushkin:2025yks}. Compared with similar mode filtering techniques used in Ref.~\cite{Dhurkunde:2026jhp}, our focus is more on the symmetric mass case ($q>0.2$), which serves as a complement to the parameter space ($q<0.2$) explored in Ref.~\cite{Dhurkunde:2026jhp}. We also cover a larger mass parameter space, with $M_{\rm tot} \in [6,400]M_{\odot}$. Furthermore, in this paper, we marginalize over the mode SNR ratio parameters instead of maximizing over them. This allows us to achieve about a $10\%$ increase in sensitivity volume.
 
Our mode-by-mode filtering approach can also be adapted to other template bank construction algorithms. Although we constructed our banks using geometric placement, the same mode-by-mode filtering applies to stochastically placed banks as in Ref.~\cite{Dhurkunde:2026jhp}. This flexibility may be especially useful in regions where geometric placement becomes unreliable, such as for binaries with anti-aligned spins or systems undergoing transitional precession, where the spin direction can flip relative to the orbital angular momentum and the precession opening angle $\theta_{\rm JL}$ varies rapidly with frequency. In such cases, the geometric placement algorithm is less efficient because the function $b(f)$ in Eq.~\eqref{eq:pres_harm} changes significantly. Stochastic placement may therefore be preferable in these regimes.

Our framework also extends naturally beyond the $\ell=2, m'=2$ sector considered here. For systems with asymmetric mass ratios, high total masses, or edge-on orientations, incorporating the $\ell=3, m'=3$ and $\ell=4, m'=4$ modes is needed in order to accurately model the system. The same mode-by-mode strategy used here at the matched filtering step can be applied to combine these harmonics at the statistic level. Moreover, eccentric binaries also admit a decomposition into a linear superposition of ``eccentric'' harmonics~\cite{Islam:2025rjl,Islam:2025llx}. The proposed filtering procedure can, in principle, also be adapted to search on that basis.

Finally, instead of using linear dimensionality reduction tools like SVD for template bank construction, we could explore nonlinear mappings using deep neural network models such as autoencoders. This approach may help uncover nonlinear relationships between parameters and enable more efficient representations through improved modeling of the latent space.

\acknowledgments

We thank Steve Fairhurst for helpful discussions.
JR acknowledges support from the Sherman Fairchild Foundation and the Jonathan M.\ Nelson Center for Collaborative Research.
MZ acknowledges support from the National Science Foundation NSF-BSF 2207583 and NSF 2209991 and the Nelson Center for Collaborative Research.
T.V acknowledges support from NSF grants 2012086 and 2309360, the Alfred P. Sloan Foundation through grant number FG-2023-20470, and the BSF through award number 2022136.

This research has made use of data, software and/or web tools obtained from the Gravitational Wave Open Science Center (\url{https://www.gw-openscience.org/}), a service of LIGO Laboratory, the LIGO Scientific Collaboration and the Virgo Collaboration. LIGO Laboratory and Advanced LIGO are funded by the United States National Science Foundation (NSF) as well as the Science and Technology Facilities Council (STFC) of the United Kingdom, the Max-Planck-Society (MPS), and the State of Niedersachsen/Germany for support of the construction of Advanced LIGO and construction and operation of the GEO600 detector. Additional support for Advanced LIGO was provided by the Australian Research Council. Virgo is funded, through the European Gravitational Observatory (EGO), by the French Centre National de Recherche Scientifique (CNRS), the Italian Istituto Nazionale di Fisica Nucleare (INFN) and the Dutch Nikhef, with contributions by institutions from Belgium, Germany, Greece, Hungary, Ireland, Japan, Monaco, Poland, Portugal, Spain.

\appendix

\section{Derivation of signal distribution}
\label{app:signal_dist}
For any fixed intrinsic parameters and sky/orientation factors, the (matched-filter) SNR amplitude scales with luminosity distance as
\begin{equation}
\label{eq:rho_D_relation}
    |\rho_k| \propto \frac{1}{D_L} ~.
\end{equation}
This is because the GW strain amplitude is proportional to $1/D_L$ and the inner product is linear in the strain. Let's also assume that the source is distributed uniformly in 3D space. Then the number distribution is given by
\begin{equation}
    P(D_L) d(D_L) = D_L^2 d(D_L) ~.
\end{equation}
After changing the variable from $D_L$ to $|\rho_k|$ according to Eq.~\eqref{eq:rho_D_relation}, we arrive at
\begin{equation}
    P(|\rho_k|) d|\rho_k| \propto |\rho_k|^{-4} d|\rho_k|
\end{equation}

\section{Bank parameter spacing dependence}
\label{app:bank_spacing}

\begin{figure}[h]
    \centering
    \includegraphics[width=1.0\linewidth]{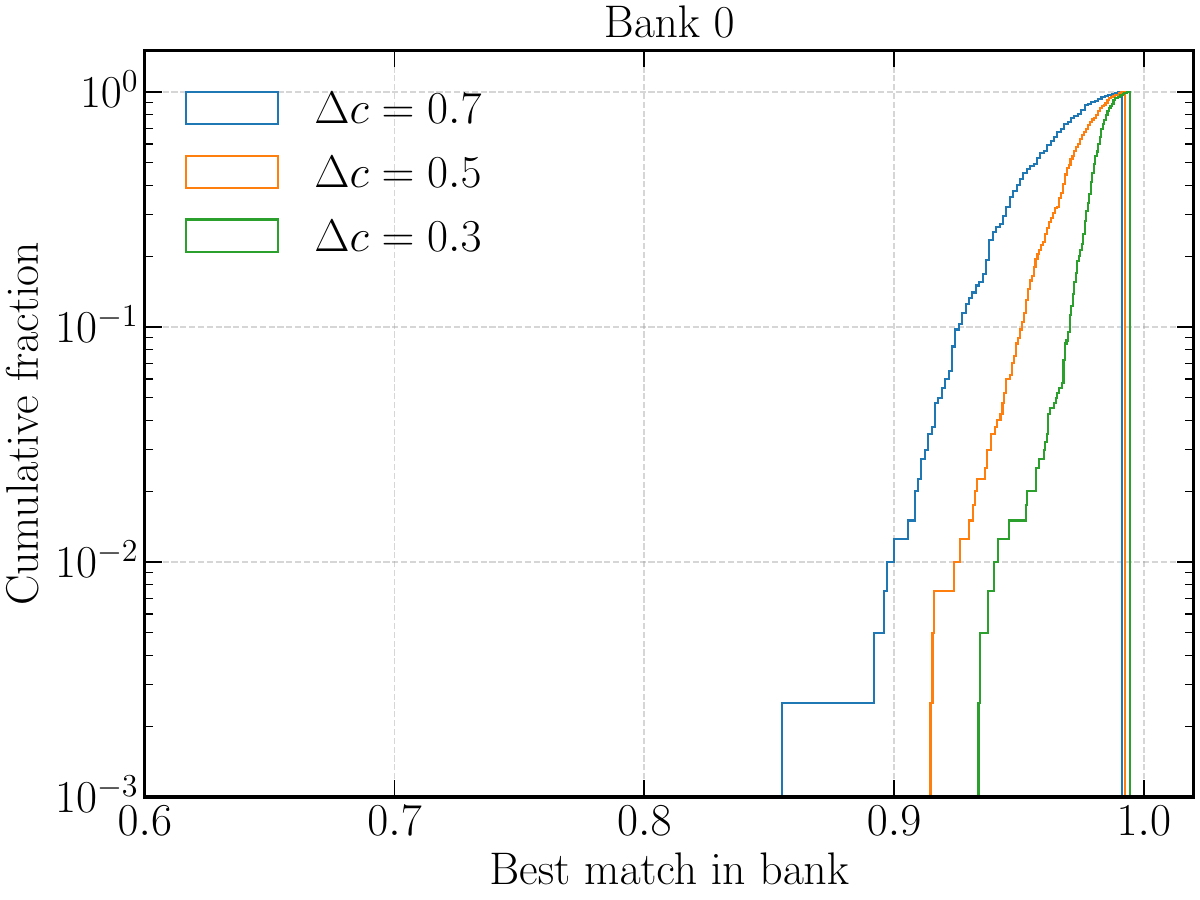}
    \caption{Effectualness of our template bank \texttt{BBH}-0 and its dependence on the parameter spacing. The total number of templates in this bank is \num{37845}, \num{86943}, \num{543026} for $\Delta c = 0.7, 0.5, 0.3$ respectively.}
    \label{fig:effectualness_bank_0}
\end{figure}

In this appendix, we show the dependence of the effectualness on the parameter spacing $\Delta \boldsymbol{c}$ within the bank. We take \texttt{BBH-0} as an example and show the effectualness for $\Delta c= 0.7, 0.5,0.3$ in Fig.~\ref{fig:effectualness_bank_0}. 
The effectualness increases by around $5\%$ percent when we shrink the spacing from $0.7$ down to $0.3$.

\clearpage
\bibliography{reference}
\end{document}